\documentstyle[11pt,aaspp4,tighten]{article} 
\lefthead{KALOGERA \& WEBBINK} 
\righthead{FORMATION OF LOW-MASS X-RAY BINARIES}
\begin{document} 
\begin{center} 
Accepted for publication in {\em The Astrophysical Journal} 
\end{center}
\title{Formation of Low-Mass X-Ray Binaries. II. 
Common Envelope Evolution of Primordial Binaries with Extreme Mass
Ratios} 
\author{Vassiliki Kalogera and Ronald F. Webbink}
\affil{Astronomy Department, University of Illinois at Urbana-Champaign,
\\ 1002 West Green St., Urbana, IL 61801. \\ e-mail: vicky,
webbink@astro.uiuc.edu} 
\newcommand{\df}{distribution function}
\newcommand{\dfs}{distribution functions} 
\newcommand{\am}{angular momentum} 
\newcommand{\os}{orbital separation}
\newcommand{\oss}{orbital separations} 
\newcommand{\br}{birth rate}
\begin{abstract} 

We study the formation of low-mass X-ray binaries (LMXBs) through helium
star supernovae in binary systems that have each emerged from a
common-envelope phase. LMXB progenitors must satisfy a large number of
evolutionary and structural constraints, including : survival through
common-envelope evolution, through the post-common-envelope phase, where
the precursor of the neutron star becomes a Wolf-Rayet star, and survival
through the supernova event. Furthermore, the binaries that survive the
explosion must reach interaction within a Hubble time and must satisfy
stability criteria for mass-transfer. These constraints, imposed under
the assumption of a symmetric supernova explosion, prohibit the formation
of short-period LMXBs transferring mass at sub-Eddington rates through
{\em any} channel in which the intermediate progenitor of the neutron
star is not completely degenerate. Barring accretion-induced collapse,
the existence of such systems therefore requires that natal kicks be
imparted to neutron stars.

We use an analytical method to synthesize the distribution of nascent
LMXBs over donor masses and orbital periods, and evaluate their birth rate
and systemic velocity dispersion. Within the limitations imposed by
observational incompleteness and selection effects, and our neglect of
secular evolution in the LMXB state, we compare our results with
observations. However, our principal objective is to evaluate how basic
model parameters (common-envelope ejection efficiency, r.m.s.\ kick
velocity, primordial mass ratio distribution) influence these results. We
conclude that the characteristics of newborn LMXBs are primarily
determined by age and stability constraints and the efficiency of magnetic
braking, and are largely independent of the primordial binary population
and the evolutionary history of LMXB progenitors (except for extreme
values of the average kick magnitude or of the common-envelope ejection
efficiency).  Theoretical estimates of total LMXB birth rates are not
credible, since they strongly depend on the observationally indeterminate
frequency of primordial binaries with extreme mass ratios in long-period
orbits.

\end{abstract}

\section{INTRODUCTION}

The existence of Low-Mass X-ray Binaries (LMXBs) poses critical questions
to the theories for the evolution of close binaries. They are believed to
be accreting neutron stars or possibly black holes with low-mass
companions (for recent reviews see Bhattacharya \& van den
Heuvel\markcite{B91} 1991; Verbunt\markcite{V93} 1993). The major problem
concerning their origin is that their orbits are now so small that they
could not accommodate the advanced evolution of the progenitor of the
compact object.  A similar question was originally posed for cataclysmic
binaries and a solution was suggested by
Paczy$\acute{\mbox{n}}$ski\markcite{P76} (1976) :  a common envelope is
formed around the binary and the spiral-in of the secondary into the
primary causes the envelope to be ejected and the orbit to contract
substantially, while exposing the degenerate core of the primary as a
newly-formed white dwarf.  A common envelope phase is adequate to solve
the puzzle of LMXBs, as well, since not only can it account for the
shrinkage of the orbit, but it also reduces the primary mass, so that the
disruptive effect of mass loss at supernova is weakened, increasing the
chance for survival of LMXB progenitors.
 
Several scenarios have been proposed for the formation of LMXBs in the
galactic disk and three out of four invoke a common-envelope phase. One
involves accretion-induced collapse (AIC) of an accreting white dwarf. The
process was first discussed by Whelan \& Iben\markcite{W73} (1973),
although in a context other than LMXB formation. A second scenario
proposes that a massive helium core, exposed in a small orbit by spiral-in
evolution, collapses to form a neutron star or a black hole (van den
Heuvel\markcite{vdH83} 1983).  A variant of this evolutionary path,
involving extensive wind mass loss in place of common-envelope evolution,
has been suggested (Romani\markcite{R92} 1992) as an avenue for producing
black-hole LMXBs.  More recently, triple-star evolution has been put
forward for LMXB formation with either a black hole or a neutron star, and
involves the formation of a Thorne-$\dot{\mbox{Z}}$ytkow star by merger of
a massive X-ray binary, and engulfment of the third component in a
common-envelope phase (Eggleton \& Verbunt\markcite{E86} 1986). A fourth
scenario has been proposed, the direct-supernova mechanism
(Kalogera\markcite{K97} 1997), which obviates the need for a common
envelope phase and relies solely on natal kicks imparted to neutron stars
to keep the systems bound and also decrease the orbital separation.

All of these scenarios present plausible formation channels for LMXBs. 
However, quantitative analysis of these evolutionary channels has been
hampered by our limited understanding of the details of the various
physical processes involved (e.g., spiral-in process, Wolf-Rayet mass
loss, asymmetric supernova explosion).  It is possible to tailor an
evolutionary model to reproduce the properties of an isolated LMXB, but
this exercise provides little perspective on whether the putative initial
conditions and subsequent tailoring are plausible. A more useful approach
is to model the evolution of an entire ensemble of primordial binaries
under a common set of assumptions, and analyze the statistical properties
of the LMXB population. Such an approach has been taken in the past for
the study of other binary populations (e.g., Lipunov \&
Postnov\markcite{L88} 1988; de Kool\markcite{K92} 1992; Kolb\markcite{K93}
1993; Tutukov \& Yungel'son\markcite{T93} 1993; Politano\markcite{P96}
1996), and more recently for LMXBs (Romani 1992; Iben, Tutukov, \&
Yungel'son\markcite{I95} 1995; Terman, Taam, \& Savage\markcite{T96}
1996).

Our purpose here is to model the evolution of a primordial binary
population through a sequence of stages involving, among others, a
common-envelope phase and the supernova explosion of a helium star, and
leading to the formation of LMXBs. Although a direct result of our
calculations if the birth frequency of LMXBs, we focus more on identifying
the properties of LMXB progenitors and on investigating the dependence of
the final population characteristics on the uncertain model parameters. We
also examine the possibility of comparing our results to observations and
constraining the observationally undetermined properties of primordial
binaries feeding LMXB formation. Although we study one evolutionary
channel here, our techniques can be straightforwardly applied to other
channels, and some of our conclusions hold for all the LMXB formation
paths that invoke a common-envelope phase. 

In \S\,2 the evolutionary scenario is described in some detail. The
relevant constraints which binaries must satisfy at various instances
throughout their evolution and the resulting limits on the LMXB-progenitor
parameter space are identified in \S\,3.  We find that asymmetric
supernova explosions are needed to explain LMXB formation via the He-star
SN mechanism, and describe the method to incorporate their effect in a
synthesis calculation in \S\,4. We discuss our assumptions for the parent
population and the synthesis method in \S\,5. The results of the
population synthesis calculations in comparison to observations as well as
their dependence on the input parameters are discussed in \S\,6. Our
conclusions are stated in \S\,7. Finally, the set of analytic
approximations employed in our model is given in an Appendix.
 
\section{DESCRIPTION OF THE EVOLUTIONARY CHANNEL}

Low-Mass X-ray Binaries have donors of mass $\lesssim 1$\,M$_\odot$.  As
elaborated below, these donors were probably always of low mass.  The
primary of a LMXB-progenitor, however, must be massive enough to produce a
neutron star. Its helium core, exposed at the end of the common-envelope
phase, must therefore have been massive enough to reach core collapse. For
these reasons, we need to consider a primordial binary system with an
extreme mass ratio.  The more massive star evolves much faster than its
companion and is the first to fill its Roche lobe.  The fact that
initially the system had an extreme mass ratio affects its evolution in
two ways: (a) The time scale for nuclear evolution of the primary is so
much smaller than that of its companion that, when mass transfer begins,
the secondary is practically still on the Zero-Age Main Sequence (ZAMS);
and (b) as its mass increases the secondary relaxes toward thermal
equilibrium on its own thermal time scale, which is long compared to the
mass transfer time scale, dictated by the thermal or dynamical time scale
of the massive donor. Consequently, the transferred material cannot cool
as it is accreted and the secondary swells up and fills its Roche lobe. 
In this way, a common envelope (CE) is created that engulfs the binary. 

Even before the formation of the common envelope, when the massive primary
approaches its Roche lobe radius, spiral-in of the secondary is initiated,
as the primary's angular momentum at synchronism exceeds one third of that
of the orbit and the Darwin tidal instability sets in (Darwin\markcite{79}
1879).  With the formation of the common envelope the secondary further
spirals toward the core of the primary due to frictional dissipation of
the orbital energy.  The details of the physical processes involved are
not well understood, but it is generally accepted that, as energy is
dissipated in the common envelope, the envelope expands and is eventually
expelled. The orbital energy is assumed to be deposited in the envelope
with an efficiency $\alpha_{CE}$ (common-envelope efficiency).  If the
orbital energy is sufficient, the binary system emerges with the secondary
and the core of the primary orbiting each other.  The post-common-envelope
orbit is considerably smaller than the initial one due to the typically
large ratio of the envelope mass to secondary mass (eq.\,[A8]). 
 
Numerical calculations of the common-envelope phase (for a review, see
Iben \& Livio\markcite{I93} 1993) show that its duration is orders of
magnitude smaller than the nuclear time scales of both the donor and the
accretor. Furthermore, Hjellming \& Taam\markcite{H91} (1991) showed that
the secondary remains practically unaffected at the end of the process and
the increase (or decrease) of its mass is insignificant ($\lesssim 1$\%).
Accordingly, we may assume that at the end of the CE phase the secondary
preserves its mass and is still on the ZAMS.  In addition, model
calculations show that, as a rule, mass transfer once started will
continue until the donor star is stripped down to a composition boundary
(Paczy$\acute{\mbox{n}}$ski\markcite{P71} 1971). We may therefore assume
that the mass of the post-CE primary is equal to the mass within its
nuclear-burning core (or within the outermost nuclear-burning shell) at
the moment it filled its Roche lobe.  In this evolutionary scenario, the
binary emerging from the common envelope evolves ``quietly'' as a detached
system until the remnant core explodes as a supernova. 

It should be noted that the binary evolution both before and after the
CE-phase is not conservative. The primaries of interest are so massive
that wind mass loss is expected to take place before the primary fills its
Roche lobe. This mass loss affects the structure and evolution of the
primary as well as the orbital characteristics of the system. Moreover,
the core of the primary emerging from the CE is still massive enough to
suffer substantial wind mass loss in a way analogous to that of a
Wolf-Rayet star. Once again the evolution of both the star and the orbit
is affected. 

The supernova explosion is a crucial event in the evolution of the
LMXB-progenitors. Most systems are disrupted, but some fraction of them
must survive if they are to evolve further to become LMXBs. We will show
later that both the survival fraction and the characteristic properties of
the newly formed systems depend strongly on the existence and mean
magnitude of a kick velocity imparted to the newborn neutron star. 

The systems that survive the supernova event can come into contact via two
physical processes: nuclear evolution of the secondary, and shrinkage of
the orbit (and hence of the Roche lobe) due to angular momentum losses.
Depending on the nature of the secondary, the physical mechanism
responsible for angular momentum losses may be gravitational radiation
and/or a magnetic stellar wind. Either way, the system comes into contact
and mass starts flowing from the secondary towards the neutron star. At
the time of contact we call the system a Zero-Age Low-Mass X-ray Binary
(ZALMXB).

\section{CONSTRAINTS AND LIMITS ON THE LMXB-PROGENITORS}

\subsection{Structural and Evolutionary Constraints}

Only a very small fraction of all binary systems follow the evolutionary
channel described above. By demanding that a system survive all
evolutionary stages in this specific sequence, we are able to constrain
the characteristics and physical parameters of the initial binaries, the
LMXB-progenitors. 

A number of constraints are imposed by this scenario (Webbink \&
Kalogera\markcite{W94} 1994) :  
\begin{enumerate} 
\item {\em The primary must fill its Roche lobe before it explodes as a
supernova.} The orbit of the progenitor cannot be arbitrarily large, since
the system must reach interaction, and enter common-envelope evolution
before the primary becomes a neutron star. \item {\em The system must
remain detached following the CE phase until the primary becomes a neutron
star.} This is a two-fold constraint: a) The orbit at the end of the
CE-phase must be wide enough to accommodate the low-mass companion; b) it
must also be wide enough not to abort evolution of the remnant core prior
to its supernova explosion. The post-CE primary is a helium star (He-star)
losing mass in a copious Wolf-Rayet (WR) wind.  Woosley, Langer, \&
Weaver\markcite{W95} (1995) have evolved mass-losing He-stars with masses
from $4$\,M$_\odot$ to $20$\,M$_\odot$, and found that they produce iron
cores barely massive enough to collapse to a neutron star. We expect that
an episode of mass transfer occurring early or midway in the evolution of
the He-star will arrest the growth of the iron core, (by completely
stripping away the helium envelope feeding it), thus preventing the
formation of a neutron star.  
\item {\em The system must remain bound after the supernova event.} Under
the assumption of a symmetric supernova, there is an absolute limit on the
amount of mass lost in the event, for the binary to survive
(Boersma\markcite{B61} 1961). If we take into account a kick velocity
imparted to the newborn neutron star due to an asymmetric core collapse,
then survival depends on the magnitude and the direction of the kick. 
\item {\em The mass transfer phase following the formation of the neutron
star must be appreciably long-lived.} In order for the system to become a
LMXB with an appreciable lifetime, the companion to the neutron star must
remain in equilibrium and the mass transfer rate must not exceed the
Eddington limit ($\dot{M}_{Edd} \sim 10^{-8}$\,M$_\odot$\,yr$^{-1}$).
However, we will entertain the possibility that a system initially
transferring mass at super-Eddington rates may find the mass transfer rate
subsiding below that limit if the companion remains in thermal and
hydrostatic equilibrium. 
\item {\em The post-SN system must reach interaction in a Hubble time.} In
order for a system to be included in the LMXB population, it must become a
luminous X-ray source within a Hubble time. This means that the post-SN
orbit must be small enough so that the secondary will fill its Roche lobe
in $\sim 10^{10}$\,yr, either due to its own evolution or due to the
shrinkage of the orbit caused by angular momentum losses. 
\end{enumerate}

\subsection{Limits on the Parameter Space of LMXB-Progenitors}

A binary system is characterized primarily by three parameters: the masses
of the two stars, $M_1$ and $M_2$ and their orbital separation $A$.
Eccentricity is another characteristic, but we will neglect it here,
assuming that tidal dissipation is efficient enough to destroy any initial
eccentricity prior to actual mass transfer. For a scale-less distribution
in orbital separation, as we will assume (\S\,5.1), the distribution of
separations of circularized orbits will be identical to that of the
initial (eccentric) orbits, so long as the distribution of eccentricities
does not itself vary significantly with separation over the range of
interest.  We can therefore assume equivalently that all the progenitors
are formed with circular orbits. The constraints described qualitatively
above substantially limit the range of values that $M_1$, $M_2$ and $A$
can cover and yet produce LMXBs.  In the calculation of these limits we
use a number of approximate relations described in detail in the Appendix. 

For specified masses\footnote{In this paper, radii and orbital separations
are expressed in terms of R$_\odot$, masses in M$_\odot$, orbital periods
in days, and time in years.} of the primary and its companion, the first
of the constraints listed above sets an upper limit on the orbital
separations of the progenitors. This limit corresponds to the primaries
that first fill their Roche lobe just before core collapse. If we choose a
value for $\alpha_{CE}$, we can find the corresponding upper limit on the
post-CE orbital separations. 

The second of the constraints sets two lower limits on the orbital
separations of the post-CE systems. One corresponds to the secondary just
filling its Roche lobe at the end of the CE phase and the other to the
He-star primary filling its Roche lobe just prior to core collapse. During
their evolution, He-stars lose mass in a strong WR wind and experience a
rapid growth in radius, which is more severe as the stellar mass decreases
(see Habets\markcite{H85} 1985;  Woosley, Langer, \& Weaver 1995).  The
radii just prior to core collapse are considerably larger than those of
the low-mass companions at ZAMS, so that the second of the constraints
obviates the first one.  The expansion of the secondary due to its own
nuclear evolution prior to the supernova is invariably negligible, since
the lifetime of the post-CE neutron star progenitor varies from $10^5$ to
$10^6$\,yr (depending on its composition at the end of the CE phase),
which is orders of magnitude smaller than the evolutionary time scale of
the low-mass companion. 

The evolutionary sequences of mass losing stars ($M < 40$\,M$_\odot$) 
presented by Schaller et al.\markcite{S92} (1992) show that massive stars
suffer most of their mass loss only during the nuclear-burning phases of
the core (H and He), when there is little or no radius expansion. In
contrast, rapid growth in radius occurs between core hydrogen exhaustion
and core helium ignition and again after helium exhaustion.  During these
phases of rapid expansion, the stellar mass is nearly constant (Figure 1).
If mass is lost to infinity from one or both components of a binary, and
carries with it a specific angular momentum equal to the orbital angular
momentum per unit mass of its source component(s), then the binary
separation varies as the inverse of the total mass of the binary (Jeans
mode of mass loss). During core He-burning slow expansion but extensive
mass loss characterizes massive stars and we find that the rate of
Roche-lobe expansion due to systemic mass loss invariably exceeds the
evolutionary rate of stellar expansion. Therefore, the primary can only
fill its Roche lobe either (i) before central He-ignition or (ii) after
central He-exhaustion. In the first case, the post-CE primary will be a
helium star with a lifetime of $\sim 10^6$\,yr (Habets 1985) losing mass
in a Wolf-Rayet wind.  These stars apparently lose most of their mass
during this phase, leading to some orbital expansion, but they also
develop denser cores and much more extended envelopes at lower masses than
would otherwise be the case. The net effect is to demand a much larger
post-CE binary separation to accommodate the evolutionary expansion of the
core He-burning primary than would be the case if it evolved at constant
mass.  In the second case, where the common envelope is formed after
central He exhaustion in the massive progenitor, the post-CE primary is
again a helium star but has a C-O core. It is also more massive (by about
1.1\,M$_\odot$) than the helium star in case (i) because
of core growth during the hydrogen-shell burning phase experienced by the
primary before CE formation. Furthermore, since helium has already been
exhausted in the center, the helium-star has also a shorter lifetime
($\sim 10^5$\,yr)  (Habets 1985), and therefore suffers minor further mass
loss, which can be ignored (Woosley, Langer, \& Weaver 1995).  Therefore
they remain massive enough so that the growth in radius is mild and hence
the limit on the \os\ is lower.  The relation between the limits is
depicted in Figure 2, from which it becomes evident that LMXB-progenitors
survive post-CE evolution up to the point of SN explosion {\em only} in
the case that the common envelope is formed {\em after} central
He-exhaustion, at which point the initial primary has already lost a
significant amount of its envelope due to its own wind.
 
In the event of a symmetric core collapse and a circular pre-SN orbit, the
system will remain bound (constraint 3) only if less than half of its
initial total mass is lost in the explosion. The assumption of a circular
orbit before the explosion is well justified, since the system has emerged
out of a common envelope, a highly dissipative process. Given a symmetric
collapse (in the frame of the primary), the binary will remain bound only
if:
\begin{eqnarray}
(M_{He}-M_{NS}) & < & \frac{M_{He}+M_2}{2}~~~~\mbox{or} \nonumber \\
M_{He} & < & M_2+2M_{NS} 
\end{eqnarray}
where $M_{He}$, $M_2$ and $M_{NS}$ are the (gravitational) masses of the
neutron star progenitor, the secondary and the neutron star respectively. 

The limits imposed on masses and radii of LMXB-donors by the final two
constraints listed above have already been studied in detail by Kalogera
\& Webbink\markcite{KW96} (1996), hereafter Paper I. Here, we summarize
their results: 

In the case of conservative mass transfer, main-sequence donors less
massive than $\sim 1.5$\,M$_\odot$ are stable against thermal time scale
mass transfer, while those crossing the Hertzsprung gap are stable if
their masses do not exceed $\sim 1.3$\,M$_\odot$. Donors that have evolved
beyond the base of the giant branch are stable against mass transfer on a
dynamical time scale and drive sub-Eddington mass transfer only if their
masses are smaller than $\sim 1$\,M$_\odot$. However, the population of
these donors is diminished by the constraint that their age must not
exceed the galactic disk age, $T$. For $T=10^{10}$\,years the parameter
space ($\log M_2$ - $\log R_2$) occupied by donors first filling their
Roche lobes beyond the base of the giant branch and transferring mass at
sub-Eddington rates is extremely small (see Figure 9a in Paper I), and
vanishes altogether if angular momentum losses due to magnetic stellar
winds are significant \footnote{Magnetic stellar wind losses were
inadvertently neglected in our estimates of initial mass transfer rates in
Paper I. Only for giant branch donors is the division between sub- and
super-Eddington systems measurably affected; none of the stability limits
is affected.}. If super-Eddington mass transfer rates are allowed, but
still with the constraint that donors remain in dynamical and thermal
equilibrium, the limits on donor masses are extended to $\sim
2$\,M$_\odot$ on the main sequence, and to $\sim 1.5$\,M$_\odot$ on the
giant branch. However, it is not clear whether these systems will actually
emerge as X-ray sources. Finally, there are two additional groups of
systems, with donors first filling their Roche lobes while on the main
sequence or while crossing the Hertzsprung gap, that experience thermal
time scale mass transfer but eventually recover equilibrium and enter a
long-lived mass transfer phase.  Those with donors filling their lobes in
the Hertzsprung gap all subside to sub-Eddington rates and emerge as
systems with giant branch donors.  However, only a portion of those with
the main-sequence donors will drive mass transfer at rates below the
Eddington limit after recovering thermal equilibrium (see Figure 6 in
Paper I). 

All relevant limits imposed on the post-CE orbital characteristics are
illustrated in Figure 3 for $M_2=1.0$\,M$_\odot$ and $\alpha_{CE} = 1$
under the assumption of a symmetric supernova. Indeed, if we adhere to the
requirement that mass transfer be sub-Eddington, we find no combination of
limits that leaves viable sub-Eddington LMXB progenitors. We conclude that
binaries could not form short-period LMXBs via this evolutionary channel
if supernovae were symmetric, regardless of the rest of their
characteristics, because the only systems which can survive mass loss in
the supernova event are so wide (in order to accommodate the evolution of
the core) that they will subsequently reach mass transfer only as the
secondary ascends the giant branch. This process will take more than
$10^{10}$\,yr (if $M_2 \lesssim 1$\,M$_\odot$), or will result in
super-Eddington mass transfer rates (if $1$\,M$_\odot \lesssim M_2
\lesssim 1.5$\,M$_\odot$), or will lead to dynamical instability (if $M_2
\gtrsim 1.5$\,M$_\odot$). The existence of short-period LMXBs therefore
demand that one or more of the constraints be relaxed.

\section{ASYMMETRIC SUPERNOVA EXPLOSIONS}

Studies of the pulsar population (e.g., Harrison, Lyne \&
Anderson\markcite{H93} 1993) show that it is characterized by a large
scale height and high space velocities, providing observational evidence
that, at their birth, pulsars are given a kick velocity, due to an
asymmetry associated with the supernova explosion. The magnitude of the
kick is large enough to influence the kinematics of the pulsar population
and certainly the orbital dynamics of a binary system hosting a neutron
star progenitor. The constraints discussed in the previous section imply
that, unless a kick velocity is imparted to the newborn compact star, it
is essentially impossible to form short-period LMXBs via the evolutionary
path considered here. Models attempting to explain the pulsar velocity
distribution and the putative velocity-magnetic moment correlation (Dewey
\& Cordes\markcite{D87} 1987; Bailes\markcite{B89} 1989) require kick
velocities with mean magnitudes of $\sim 100-200$\,km\,s$^{-1}$. However,
a more recent study (Lyne \&\ Lorimer\markcite{L94} 1994) of the pulsar
population takes into account a selection effect against high velocity
pulsars, and concludes that the mean pulsar velocity is $\sim
450$\,km\,s$^{-1}$. Additional evidence from supernova remnants and
associated pulsar positions (Caraveo\markcite{C93} 1993; Frail, Goss, \&
Whiteoak\markcite{F94} 1994) supports the conclusion of high kick
velocities. Although pulsar velocities do not directly reflect the birth
velocities, these recent estimates do point towards high kick magnitudes.
Any correlation between kick direction or magnitude and orbital axis or
orbital velocity in a binary is at present purely conjectural, and hence
we will assume that kick velocities are isotropically oriented in the
center of mass frame of the collapsing component with a Maxwellian
distribution in magnitude. 

The interplay between the different limits discussed in the previous
section changes dramatically if we relax the assumption of a symmetric
supernova explosion. An asymmetric core collapse, imparting a kick
velocity to the neutron star, breaks the one-to-one link between pre- and
post-SN orbital parameters.  Those constraints in Figure 3 which reflect
post-SN conditions no longer sharply delimit possible LMXB progenitors. 
Systems which in the case of symmetric supernovae would have certainly
been disrupted may now survive (if by chance the kick velocity has the
right direction and magnitude), and, conversely, systems which would have
survived may now be disrupted. Moreover, post-supernova orbits may now
become smaller than the pre-supernova ones (which can never be the case in
a symmetric core collapse), allowing the formation of short-period LMXBs.
Thus, for the case of an asymmetric collapse the limits imposed on the
progenitors, after the ejection of the common envelope, are only the ones
shown in Figure 4. In that case, a non-vanishing part of the parameter
space may be populated by LMXB progenitors. The post-CE progenitors are
Wolf-Rayet binaries, and for a $1$\,M$_\odot$ secondary they have
primaries with masses $\sim 3.5 - 8$\,M$_\odot$, orbital separations $\sim
8 - 25$\,R$_\odot$, and orbital periods $\sim 1 - 5$\,d.  The
corresponding limits on the primordial binaries are also shown in Figure
4; these O,B primaries have masses $\sim 13 - 25$\,M$_\odot$, orbital
separations $\sim 800 - 1800$\,R$_\odot$, and orbital periods $\sim 1.5 -
5$\,yr.

The inclusion of a kick velocity imparted to the neutron star forces one
to follow the evolution of an initial population of binaries and not of a
single system.  The stochastic element in this problem, of finding the
distribution of binaries after an asymmetric supernova explosion, has been
already addressed by Kalogera\markcite{K96} (1996). Assuming an
isotropic Maxwellian distribution of kick velocities, she developed an
analytical method of calculating the distribution of post-SN binary
systems over eccentricity, orbital separations (before and after
circularization) and systemic velocities. Here, we are interested only in
the distribution of orbital separations of post-SN circularized orbits.
Following the notation of Kalogera (1996), the distribution of systems
over of the dimensionless separation $\alpha_c \equiv A_c/A_i$, where
$A_c$ and $A_i$ are the circularized and pre-SN orbital separations,
respectively, is given by: 
\begin{equation} {\cal H}(\alpha_c)~=~
\left(\frac{\beta}{2\xi^2}\right)~ \exp\left(\frac{-(\beta
\alpha_c+1)}{2\xi^2}\right)~I_o\left(\frac{\sqrt{\beta\alpha_c}}{\xi^2}
\right)~\mbox{erf}\left(z_o\sqrt{\frac{\beta}{2\xi^2}}\right),
\end{equation} 
where 
\begin{eqnarray} \mbox{erf}(x_o) & \equiv &
\frac{2}{\sqrt{\pi}} \int_0^{x_o}\mbox{e}^{-x^2}~ dx\mbox{,} \nonumber \\
z_o & = & \sqrt{2-\alpha_c-\frac{2c-\alpha_c}{c^2}}\mbox{,}~~~~~~~~~
\frac{2c}{1+c}<\alpha_c<2c \nonumber \\
 & = & \sqrt{2-\alpha_c}\mbox{,}\hspace{3.7cm} 2c\leq\alpha_c<2 \nonumber \\
 \beta & = & \frac{M_{NS}+M_2}{M_c+M_2}, \nonumber \\
\xi & = & \frac{\sigma}{V_r}, \nonumber  
\end{eqnarray}
$I_o$ is the zeroth order Bessel function, $\sigma = \langle
V_k^2/3\rangle ^{1/2}$, $V_r$ is the relative orbital velocity of the two
stars in the pre-SN binary, and $c$ is the ratio of the radius of the
secondary to the pre-SN orbital separation. 

Convolving the above distribution with that of the pre-SN binaries over
masses and orbital separations, as defined by the limits already
discussed, enables us to map precisely the distribution of post-SN
binaries and synthesize the population of nascent LMXBs.

\section{POPULATION SYNTHESIS}

\subsection{Parent Binary Evolution}

Having described the criteria which select LMXB progenitors from a parent
binary population, we require a statistical description of this primordial
population to produce quantitative results.  We therefore assume that the
primordial binaries can be characterized by three parameters : the mass of
the primary $M_1$, the mass ratio $q\equiv M_2/M_1$ ($M_2$ being the mass
of the secondary star), and the orbital separation of the system $A$. In
selecting an initial distribution of binaries over these parameters, we
are guided by the results of a detailed analysis by Hogeveen\markcite{H91}
(1991), but with some important differences at small mass ratios, where
observational constraints are virtually non-existent.
 
We have adopted the field star Initial Mass Function (IMF) derived by
Scalo\markcite{S86} (1986) as a good representation of the primary mass
distribution.  Based on his results we are able to fit the IMF of stars
more massive than $0.3$\,M$_\odot$ with a single power law of the form : 
\begin{equation}
\Xi(M) = \Xi_o M^{-2.7}~~\mbox{stars}~\mbox{pc}^{-2}\mbox{yr}^{-1}\mbox{M}_\odot ^{-1},~~~\Xi_o\simeq 6.83\times 10^{-10}
\end{equation}
If we assume that the galactic disk has an exponential surface density
with a scale length of $4$\,kpc, and that the distance of the Sun from the
galactic center is $8$\,kpc, then we estimate the effective radius of the
galactic disk to be $15$\,kpc. The birth rate of primaries per unit
logarithm of mass, integrated over the entire galactic disk, is then : 
\begin{equation}
f_1(\log M_1) \simeq 1.112~M_1^{-1.7}~\mbox{yr}^{-1}~
(\log \mbox{M}_\odot)^{-1}
\end{equation}

The distribution of orbital separations is assumed to be inversely
proportional to A (Abt\markcite{A83} 1983), normalized to a wide range of
initial separations up to $10^6$\,R$_\odot$. This assumption may appear
inconsistent with more recent results regarding the orbital period
distribution of solar-type binaries (Duquennoy \& Mayor\markcite{D91}
1991). However, we note that the range of orbital separations, hence
orbital periods, of interest to us is extremely narrow, from $\sim 2$\,yr
to $\sim 5$\,yr, so that our results are not sensitive to the specific
shape of the broader distribution.  Furthermore, our choice of the
functional form and normalization is consistent with the one used by
Hogeveen (1991) in his study of the mass ratio distribution, the results
of which we have chosen to adopt. 

The mass ratio distribution of unevolved binaries of interest to us is
quite uncertain. It is empirically known only in the limit of
approximately equal component masses and for relatively close binaries.
Results obtained by Hogeveen (1991) show that for $q\gtrsim 0.35$ the mass
ratio distribution at long orbital periods is described by an IMF-like
power law ($\propto q^{-2.7}$). However, we need to extrapolate to very
small values of $q$ ($< 0.1$). For this range of values it is often
assumed that the distribution flattens, but this is in truth an ad hoc
assumption, because the contribution of such extreme mass ratio systems to
the observed distribution of spectroscopic or eclipsing binaries at long
periods ($>1$\,yr) is negligible.  Instead we have chosen to adopt an
IMF-like $q$-distribution, even for very small values of $q$.  By making
this assumption, and demanding that the normalization accords with
observation as $q \rightarrow 1$, we must explicitly allow for the
possibility that our primordial systems are not only binary, but multiple.
In doing so, we recognize that the presence of additional stellar
components modifies our pool of progenitor binaries in two ways : (i) an
inner binary may abort evolution of the primary by mass exchange,
thwarting its expansion to a common-envelope stage involving the secondary
component of interest to our scenario; and (ii) triple systems are
dynamically stable only if the period ratio between outer and inner orbits
exceeds some critical value. Regarding the first of these two elements, an
inner binary with a secondary component {\it less} massive than the outer
one of interest to us is very unlikely to be of any consequence : the
inner binary will succumb to common-envelope evolution, but it is
incapable of extracting enough energy to eject the envelope before merging
-- the outer binaries of interest to us typically only barely manage to
survive. We therefore exclude from our progenitor pool only those
multiples in which the inner binary contains a more massive secondary than
the outer. Similarly, in regard to the second element, dynamical
instability of a triple star typically leads to ejection of the least
massive component (Harrington\markcite{H75} 1975). We therefore exclude
from our progenitor pool only those multiples in which a third component,
more massive than the secondary of interest to us, lies within a critical
period (or separation) ratio of the secondary orbit. Following Kiseleva,
Eggleton, \& Anosova\markcite{K94} (1994), we adopt a critical period
ratio of $6.3$ (separation ratio $\simeq 3.4$) for the extreme mass ratios
of interest here. All systems containing third components more massive
than our secondary are therefore excluded, from a maximum orbital period
of $6.3$ times that of interest down to a minimum physically allowable
separation, which we take (for simplicity) to be twice the primary radius.
Assuming that binary and multiple stars are chosen from a parent
population according to Poisson statistics (i.e., that they are
independent, uncorrelated events), we modify our simple inverse
distribution in $A$ and power-law distribution in $q$ by a factor
representing the Poisson probability that neither of the above
strictures is violated : 
\begin{equation}
g(q,A)~=~\frac{0.075}{A}~0.04q^{-2.7}~\exp \left( - \int_{2R(M_1)}^{A\cdot 6.3^
{2/3}}\int_{q}^{1}~0.075A'^{-1}~0.04q'^{-2.7}~dA'dq' \right).
\end{equation}
A plot of this assumed distribution over mass ratio, $q$, for specified
primary mass, $M_1$, and orbital separation, $A$, is shown in Figure 5. It
bears re-emphasizing that this distribution is unverifiable by current
observation for $q \lesssim 0.35$. The adoption of equation (5) is
motivated by three factors : (1) it is consistent with observed rates of
duplicity and mass ratio, where these are detectable, for binary
separations of interest to us;  (2) it is a logical extrapolation of that
observable part of the distribution to the extreme mass ratios of interest
to us, without the invocation of ad hoc breaks or cut-offs; and (3) it
provides a consistent formalism for future modeling of LMXB formation by
triple star evolution. 

We can transform equation (5) to a distribution over $\log M_2$ and $\log
A$, $h_{in}(\log M_2,\log A)$, using the definition of $q$.  The
distribution representing the primordial binary population then becomes : 
\begin{equation}
F_{in}(\log M_1,\log M_2,\log A)~=~f_1(\log M_1)\cdot h_{in}(\log M_2,\log A)
\end{equation}
The range of values covered by the three parameters is dictated by the
evolutionary selection criteria already discussed. 

\subsection{Method}

Having defined the parent binary population, we are able to follow its
transformation as the systems evolve through the various evolutionary
stages.  This is done by identifying the system parameters at the end of
each stage and their dependence on the corresponding parameters at the
beginning of each phase, and by transforming the distribution function
according to these dependences.  These transformations are performed
analytically, so that at each stage prior to the explosion the
distribution function of binaries can be expressed explicitly. At the
supernova stage the pre-SN function is convolved with the distribution
over post-SN circularized separations (eq.\,[2]), and the product is
integrated numerically now over pre-SN helium-star masses and orbital
separations. This method offers major advantages over Monte Carlo
techniques as it is free of any statistical errors and in principle allows
us to have an infinite resolution in the final LMXB parameters. This high
resolution reveals even the most subtle features in the nascent LMXB
distribution and permits us to trace back the origin of these features. In
what follows, we briefly describe the procedure for each evolutionary
stage of interest. 

From all the systems represented by $F_{in}$, we are interested only in
those that experience a common-envelope phase. The post-CE systems are
characterized by the secondary mass $M_2$ (assumed unchanged by CE
evolution), the orbital separation $A_{post-CE}$, and the mass of the
remnant core $M_{He}$, which depends only on the primary mass. Using the
relations connecting the pre- and post-CE binary parameters we can find
analytically the transformed post-CE distribution function :
\begin{equation}
F_{CE}(\log M_{He},\log M_2,\log A_{post-CE})  =  F_{in}~\cdot ~ 
J\left(\frac{\log M_1,\log M_2,\log A}{\log M_{He},\log M_2,\log A_{post-CE}}\right). 
\end{equation}
Since $\partial \log A/\partial \log A_{post-CE}=1$ (eq.\,[A8]), $M_2$ is
unchanged, and $M_{He}$ is a function only of $M_1$ (eq.\,[A3]), the
distribution of post-CE \oss\ and secondary masses for a specific choice
of $M_{He}$ is simply a homologous transformation of their pre-CE
distribution at the corresponding value of $M_1$. 

The post-CE primary, $M_{He}$, has already exhausted helium in its core,
since the initial primary entered common-envelope evolution after core-He
exhaustion. The time scale for nuclear evolution of the C-O core until
collapse is $\sim 10^5$\,yr (Habets 1985), and is so short that the helium
star is essentially unaffected by wind mass loss (Woosley, Langer, \&
Weaver 1995). Therefore the pre-SN distribution of binaries is identical
with the one just after the CE phase.  The secondary is still on the main
sequence when the supernova occurs.

By convolving the pre-SN distribution with the survival probability
distribution for the supernova explosion, ${\cal H}(\alpha_c)$ (eq.\,[2]),
we can obtain the \df , $Z(\log M_2,\log A_{post-SN})$, of post-SN
circularized \oss\ $A_{post-SN}$ and secondary masses $M_2$ by integrating
over $M_{He}$ and $A_{pre-SN}$. In performing this transformation, we
assume that all He stars leave a remnant neutron star of the same
gravitational mass (see also Woosley, Langer, \& Weaver 1995) of
$1.4$\,M$_\odot$.  The post-SN distribution thus becomes a two-variable
function of $M_2$ and $A_{post-SN}$:
\begin{equation}
Z(\log M_2,\log A_{post-SN}) =
\int_{\log M_{He}^{min}}^{\log M_{He}^{max}}~\int_
{\log A_{pre-SN}^{min}}^{\log A_{pre-SN}^{max}}
\zeta~d\log A_{pre-SN}~d\log M_{He}, 
\end{equation}
where 
\begin{displaymath}
\zeta \equiv F_{CE}\cdot {\cal H}(\alpha_c)\cdot 
\alpha_c \ln  10~,
\end{displaymath}
and $\alpha_c \ln 10$ is the Jacobian corresponding to the variable 
transformation from 
$\alpha_c = A_{post-SN}/A_{pre-SN}$ to $\log A_{post-SN}$.
The limits of the integration over $\log A_{pre-SN}$ depend on both $M_{He}$ 
and $M_2$; those for   
the integration over $M_{He}$ depend on $M_2$, according to the constraints 
discussed in \S\,3.

We have assumed here that both synchronization and circularization of
the binary occurs relatively soon and certainly prior to the time the
secondary overflows its Roche lobe. The assumption is well justified
since the time scales for both processes for detached systems are
significantly shorter than the evolutionary time scale of the secondary
as well as the time scale for angular momentum losses due to magnetic
braking. As the binary approaches Roche lobe overflow the time scales
rapidly decrease down to tens to thousands of years (e.g., for
$R_L/R_2\simeq 2$; see Zahn\markcite{Z77}\markcite{Z89} 1977, 1989).

Systems surviving the supernova event do not all form LMXBs.  Binaries
must still evolve further towards Roche lobe overflow of the secondary for
mass transfer to be initiated. At this stage binary evolution is driven by
nuclear evolution of the secondary and loss of \am\ , and hence shrinkage
of the Roche lobe around the secondary.  We consider two mechanisms
responsible for the loss of \am : gravitational radiation (eq.\,[A11]) and
magnetic braking (eq.\,[A13]).  In the latter process, a wind from the
secondary, locked onto the stellar magnetic field, drives \am\ away from
the star. Assuming that the companion is maintained in synchronization
with the orbit by tidal dissipation, it follows that the binary loses \am
 (Verbunt \& Zwaan\markcite{V81} 1981).  This \am\ loss affects the
orbital
characteristics considerably, whereas the mass loss rate is assumed
negligible. For very low-mass secondaries ($M_2\leq 0.37$\,M$_\odot$) 
that are fully convective, we assume that magnetic braking is negligible,
in accordance with arguments advanced to explain the $2^{\mbox{h}}-
3^{\mbox{h}}$ gap in the orbital period distribution of cataclysmic
variables (Rappaport, Verbunt \& Joss\markcite{R83} 1983). For these
masses, \am\ loss due to gravitational radiation alone is considered. 

It should be noted that studies of the magnetic braking mechanism rely
upon measurements of rotational velocities of solar-type stars (Verbunt \&
Zwaan 1981). More massive stars develop radiative envelopes which are
expected to diminish the dynamo generation of magnetic fields and hence
the effect of magnetic braking. In accordance to this, massive stars
appear to rotate much faster than low-mass stars.
We have adopted the functional form used by Rappaport et al.\ (1983)
(with their index $\gamma =2$), but modifying the braking efficiency for
stars more massive than the Sun by introducing a cutoff factor, $b$,
dependent only on stellar mass.
Using observed mean rotational
velocities for main sequence stars, we were able to estimate the efficiency
factor,
$b(M_2)$:
\begin{eqnarray}
b(M_2) & = & 0
~~~~~~~~~~~~~~~~~~~~~~~~~~~~~~~~~~~~~~~~~~~~~~~~~~~~~~~M_2 \leq
0.37\mbox{\,M\,}_\odot, \nonumber \\
& = & 1
~~~~~~~~~~~~~~~~~~~~~~~~~~~~~~~~~~~~~~~~ 0.37\mbox{\,M\,}_\odot < M_2 \leq
1.03\mbox{\,M\,}_\odot, \nonumber \\
& =
&\mbox{exp}~\left[-4.15~(M_2-1.03)\right]~~~~~~~~~~~~~~~~~~~~~~~~M_2 >
1.03\mbox{\,M\,}_\odot.
\end{eqnarray}
This expression for the magnetic braking efficiency reproduces the
rotation velocities of main sequence stars of spectral types F5 and F0
(Allen 1973) assuming that they are born at rotational
break-up and neglecting evolutionary changes in mass and radius. Main
sequence stars of earlier spectral type show no evidence of magnetic
braking. Using more recent data (e.g., Fukuda\markcite{F82} 1982; 
Kawaler\markcite{K87} 1987) leads to somewhat different expressions for 
$b(M_2)$, but has no qualitative effect on our results.  
Because of the assumption of initial maximum rotation the above
estimate is actually an {\em upper} limit to the magnetic braking efficiency
factor.

The last step in evolving the \df\ $Z$ is to transform the post-SN systems
to nascent LMXBs. We set the radius of the secondary (eq.\,[A9] in Paper
I) equal to its Roche lobe radius (eq.\,[A7]) and eliminate the time by
using either equation (A12) or equation (A14). The resulting equation can
be solved numerically for the orbital separation, $A_{X}$, at the onset of
the mass transfer phase.  In this way we are able to find the distribution
over orbital separation, $A_X$, and donor mass, $M_2$, of the LMXB
progenitors: 
\begin{equation}
\Phi _A(\log M_2,\log A_{X}) = Z~\cdot ~
\left\vert\frac{\partial \log A_{post-SN}}{\partial \log A_{X}}\right\vert
\end{equation}
The derivative in the above equation is calculated analytically.  With one
last transformation we obtain the distribution over donor mass and orbital
period, $\Phi _P(\log M_2,\log P_{X})$. 

\section{RESULTS}

\subsection{A Reference Model}

Results from our population synthesis calculations are illustrated in
Figures 6a and 6b for a prototypical choice of input parameters, which
we shall deem our reference case.  The two frames of this figure show
zero-age LMXB distributions, $\Phi(\log M_2, \log P_X)$, for systems
initiating sub-Eddington mass transfer only (Figure 6a), and for both
sub-Eddington and super-Eddington systems (Figure 6b).  The constraints
delineating these regions were discussed in Paper I, and are illustrated
again here in Figure 7, where the regions are labeled $S_E$ and $S^E$,
respectively.  Our choices of values for free parameters in this
reference case have been made in such a way as (i) to define a plausible
extreme, or (ii) to characterize the model distribution at the threshold
value of a specific parameter, that is, at a value where its influence
on the resulting models changes character.  Thus, for example, our
choice of mass ratio distribution (eq.\,[5]) defines a plausible upper
limit to the frequency of the massive binaries with extreme mass ratios
which feed our evolutionary channel, since the Poisson cutoff invoked
in equation (5) (an upper limit to the number of close companions a
massive star may accommodate within the limits of dynamical stability)
is taking effect in just the range of companion masses of interest (see
Figure 5).  For the common envelope ejection efficiency we choose
$\alpha_{CE}=0.3$, because below this value the survival window (the
region bounded by thick and thin solid lines in Figure 3) disappears
rapidly below the lower limits to post-supernova binary separation
imposed by the need to accommodate both the helium-star core of the
primary (the dotted line in Figure 3)  and its companion (the thin
dashed line in Figure 3).  Our choice of r.m.s. kick velocity for the
reference case, $\langle V_k^2\rangle ^{1/2}=300$\,km\,s$^{-1}$, equates
approximately to the maximum pre-SN relative orbital velocities, and
therefore lies very near the peak in their survival probability in the
zero-age LMXB population. 

Within the age and stability limits set by Figure 7, the general
features seen in Figures 6a and 6b are the result primarily of a
competition between nuclear evolution of the donor stars and angular
momentum loss from the binary.  The prominent ridge extending towards
low companion masses and low orbital periods is due to systems with
essentially zero-age donors, brought to Roche lobe contact due to loss
of angular momentum.  This ridge along the ZAMS disappears for donors
more massive than $\sim 1.4$\,M$_\odot$, because at these masses angular
momentum losses due to magnetic braking become inefficient (eq.\,[9]). 
For donors more massive than $\sim 1$\,M$_\odot$, nuclear evolution
becomes increasingly important, and not all post-SN systems experience
orbital shrinkage.  As a result, a minimum appears in the distribution
at orbital periods of about one day.  Systems with donors on the giant
branch appear only in the super- Eddington population.  They form the
broad peak at long periods between donor masses $\sim 1$\,M$_\odot$ and
$\sim 1.5$\,M$_\odot$, and have reached contact because of the advanced
nuclear evolution of the donor. 

The competition between angular momentum losses and nuclear evolution is also
evident in the distribution over orbital periods, $\Psi_P(\log P_X)$,
obtained by integrating $\Phi_P$ over $\log M_2$, and plotted in Figure 8. 
The first peak at $\sim 0.3^d$ arises from the peak in the mass ratio
distribution (cf. Figure 5), whereas the peak at $\sim 0.5^d$ is the result
of the flattening of the ZAMS radius-mass relation above $\sim
1.3$\,M$_\odot$, which compresses a relatively wide range of donor masses
into a narrow range of periods. The valley at $\sim 1^d$ is a result of
magnetic braking evacuating this range.  Systems with evolved donors that
transfer mass at super-Eddington rates populate the "bump" at longer periods. 
These systems may not at first appear as luminous X-ray sources, as we
anticipate that their dense super-Eddington outflows will quench X-ray
emission.  Nevertheless, as the donor mass decreases, mass transfer may
subside to sub-Eddington rates, and the systems will then appear as LMXBs
with donors on the giant branch.

We note in passing that Figure 8 also bears witness to the power of the
analytical technique used for these synthesis calculations to reveal features
which are very difficult and computationally expensive to identify in Monte
Carlo approaches. A case in point is the inflection point visible at $\sim
0.23^d$, below the shortest-period maximum.  This feature is in fact an
artifact of the ZAMS radius-mass relation we have adopted in this work
(eq.\,[A1] in Paper I), which is discontinuous in its first derivative at
$M_2 \simeq 0.8$\,M$_\odot$. With an analytic approach, we have the power in
principle to increase resolution within a limited range of parameter space,
as desired, without being obliged to do so everywhere, and without suffering
the Poisson noise inherent in Monte Carlo calculations.

\subsection{Observable Properties of the LMXB Population}

Despite three decades' effort in X-ray astronomy, our knowledge of the
underlying structural properties of LMXBs is still extremely limited and
fragmentary.  Orbital periods, for example, are known only for a small
minority of systems, a large fraction of LMXBs lack optical counterparts
(because of low intrinsic optical luminosity and heavy interstellar
extinction), and dynamical mass estimates from spectroscopic orbits are
nearly absent outside that collection of soft X-ray transients which
evidently contain black hole accretors of mass $>3$\,M$_\odot$ (and
which cannot originate through the formation channel modeled here).
Nevertheless, there are several bases, summarized here in Table 1, on
which a comparison may be made between global observational properties
and the results of population synthesis models.  The origin of the
observational estimates contained in Table 1 is described below;
theoretical estimates are listed separately for those systems which
transfer mass initially at sub-Eddington rates (regions $S_E$, which we
expect to remain LMXBs throughout this phase of interaction) and those
initially super-Eddington (regions $S^E$, which we expect to contribute
to the observed LMXB population only later during interaction, if at
all).  It must be emphasized here that the values of free parameters
defining our reference model, from which results are extracted in Table
1, were chosen to aid in characterizing the dependence of model results
on those parameters;  they have not been chosen to optimize agreement
between model and observation.  The reader may glean some sense of the
adjustments required from the discussion of parameter dependences which
will follow below.

Some explanations are warranted for the entries in Table 1:

{\em Birth rate}.  We estimate the birth rate of the observed population from
the catalogs of galactic LMXBs by van Paradijs (1995) and Bradt \& McClintock
(1993).  Black hole candidates and LMXBs in globular clusters have been
excluded. Distance estimates and mean X-ray luminosities of individual
systems were drawn, where available, from those catalogs. The birth rate in
steady state then follows from summing the observed mean X-ray luminosities,
and dividing by an average initial donor star mass (assumed to be $1.2$\,M$_\odot$,
as suggested by the synthesis results), and assuming an X-ray production rate
of $1.86\times 10^{20}$\,erg\,g$^{-1}$ of accreted matter. The theoretical
birth rates quoted here exclude any contribution from possible LMXB
progenitors which may emerge from thermal time scale mass transfer, regions
MS$_2$ and HG$_2$ in Figure 5 of Paper I;  the birth rates for their
immediate progenitors are, respectively, $2\times 10^{-6}$\,yr$^{-1}$ for
region MS$_1$ and $1\times 10^{-6}$\,yr$^{-1}$ for region MS$_2$, in our
reference model.

{\em Total X-ray luminosity}.  For comparison, we also include in Table 1
estimates of the observed and theoretical total X-ray luminosity for Galactic
disk LMXBs.  We derive a statistical (Poisson) uncertainty in the observed
luminosity of $\pm 30\%$, but expect the true uncertainty to be substantially
greater due to systematic errors (from spectral fittings and distance
estimate errors).  Since the deduced estimate of the birth rate of observed
LMXBs follows directly from their total X-ray luminosity, this entry does not
in reality provide a new benchmark for comparison, but it does strip away
some of the assumptions applied above to deduce an observed birth rate.  We
apply the same assumptions instead to the synthesis models to convert birth
rates to total X-ray luminosity, but now employ the actual donor mass
distribution produced by those models, instead of an average value.

{\em Fraction of short-period systems}. Secular evolution among LMXBs
produces a natural bifurcation in their evolution, with short-period systems
($P_X \lesssim 20^{\mbox{h}}$) driven to shorter orbital periods by angular
momentum loss, and long-period systems driven to longer periods by nuclear
evolution of the donor star (Taam, Flannery \& Faulkner 1980;  Pylyser \&
Savonije 1989).  This behavior provides a basis for comparison between theory
and observation, even though our synthesis models do not address secular
evolution in the LMXB state.  Unfortunately, orbital periods are known for
only 30\% of galactic LMXBs;  of the 24 systems with known periods, 18 fall
into the short-period group.  The observational upper limit quoted in Table 1
reflects our expectation that the higher optical/infrared luminosities of
donors in longer-period systems favor detection of their orbital periods, so
that LMXBs with undetected periods are more likely to belong to the
short-period group. It is important to note as well that the theoretical
estimates listed for our reference case are probably lower limits, in that
they reflect relative birth rates of short- and long-period systems, and do
not account for the shorter lifetimes expected among longer orbital period
systems.

{\em Fraction of neutron star accretors}.  A significant fraction of the
neutron stars in our model populations (at least among those transferring
mass at sub-Eddington rates) may be driven to gravitational collapse during
their X-ray lifetime, and become stellar black holes.  An observational lower
limit to the fraction of LMXBs containing neutron stars, quoted in Table 1 is
set by those showing X-ray pulsations or classical X-ray bursts (see van
Paradijs 1995).  To obtain a theoretical estimate for this fraction, we adopt
the equation of state (AV14/UVII) developed by Wiringa, Fiks \& Fabrocini
(1988), which represents the most complete microscopic calculations available
at present; this equation of state predicts maximum gravitational and
baryonic (non-rotating) neutron star masses of 2.13\,M$_\odot$ and
2.64\,M$_\odot$, respectively (Cook, Shapiro \& Teukolsky 1994).  Model
systems with total baryonic mass exceeding 2.64\,M$_\odot$ are considered to
contain black hole accretors only once the accretor mass passes that
threshold.

We must emphasize that black hole formation through accretion-induced neutron
star collapse is incapable of explaining the existence of the low-mass
black-hole soft X-ray transients A 0620-00 (V616 Mon), GS 2023+338 (V404
Cyg), GS 1124-684 (GU Mus), GRO J1655-40, GS 2000+25 (QZ Vul), and H 1705-250
(V2107 Oph) (Cowley\markcite{C94} 1994; Bailyn et al.\markcite{B95} 1995; 
Charles \& Casares\markcite{C95} 1995;  Remillard et al.\markcite{R96} 1996). 
In each of these systems, lower limits to the masses of their compact
components, derived dynamically from the reflex orbital motion of their donor
stars, clearly exceed the maximum total mass of any of our modeled systems: 
1.4\,M$_\odot + 1.5$\,M$_\odot = 2.9$\,M$_\odot$.  At least one other
evolutionary channel is required (Eggleton \& Verbunt 1986; Romani 1992).

{\em Systemic velocities}.  We have derived an observed velocity dispersion
from the tabulation by Johnston (1992) of heliocentric radial velocities of
15 LMXBs, correcting for solar motion and for differential galactic rotation,
using her distance estimates and the galactic rotation model of
Clemens\markcite{C85} (1985), and assuming isotropic peculiar velocities with
respect to uniform rotation on cylinders. Neither the rotation model nor the
assumption of isotropic peculiar velocities can be strictly valid, but the
observed velocity dispersion is more seriously suspect because of distance
errors, since differential rotation corrections are large, and because of
small-number statistics. The theoretical velocity dispersions in Table 1
reflect one-dimensional peculiar velocities at birth; virialization within
the galactic potential should reduce them by a factor of $\sqrt{2}$, since
the hiatus between supernova explosion and the onset of mass transfer as an
LMXB significantly exceeds a galactic dynamical time scale for the
overwhelming majority of model systems.

\subsection{Parameter Studies}

Although one should treat the observed quantities listed in Table 1 with some
caution, for reasons outlined above, it is instructive to explore how the
theoretical quantities listed there respond to variations in the principal
input parameters to our population models:  (i) the efficiency of common
envelope ejection, $\alpha_{CE}$; (ii) the r.m.s.  kick velocity imparted to
a newborn neutron star, $\langle 
V_k^2\rangle ^{1/2}$; (iii) the initial mass
ratio distribution, and (iv) the maximum neutron star mass. These
dependencies are summarized semi-quantitatively in Table 2, and discussed
physically below.

{\em Common envelope efficiency}.  As illustrated in Figure 3,
progenitor systems of given donor star mass populate only a narrow range
of post-common-envelope orbital separations.  That range shifts to
smaller separations for smaller companion masses (less orbital energy
available for envelope ejection) or for small ejections efficiencies,
$\alpha_{CE}$ (less efficient use of available orbital energy).  Since
the lower limits to binary separations are fixed by Roche lobe
constraints, reductions in $\alpha_{CE}$ therefore result in (i) 
progressive loss of the lowest-mass companions from the pool of donor
stars, and (ii)  progressive loss of the longest-period component of the
survivor pool.  The loss of low-mass donors suppresses the short-period
extreme of the LMXB orbital period distribution. Likewise, since
asymmetric supernovae cannot produce circularized post-supernova
separations exceeding twice the pre-supernova separation (Kalogera
1996), small values of $\alpha_{CE}$ also suppress the long-period
extreme in this distribution (see Figure 10). For $\alpha_{CE} \lesssim
0.3$, the peak of the donor mass distribution no longer survives, and
the birth rate falls precipitously (Figure 9).  The slow increase in
systemic velocity dispersion of survivors as $\alpha_{CE}$ decreases
reflects (i) the selection of survivor systems, crudely, according to
whether the supernova kick by chance imparts to the neutron star a space
velocity closely matching the orbital velocity its companion at the
instant of the explosion, and (ii) the closing of the window in
separation spanned by companion stars of different masses. 

{\em Average kick velocity}.  The dynamical consequences of supernova
kicks are described in some detail by Kalogera (1996). Aside from a nearly
uniform suppression of survival probabilities, r.m.s.\ kick velocities
exceeding the largest pre-supernova relative orbital velocities ($\sim
300$\,km\,s$^{-1}$) exercise very little influence on either the mass- and
orbital period-distribution of survivors, or on their space velocities,
since survivors then come only from the low-velocity tail of the
Maxwellian kick distribution.  However, when kick velocities are small,
they are capable only of binding relatively wide systems, which have
correspondingly small pre-supernova relative orbital velocities, and
consequently acquire only small post- supernova space velocities. (These
wide systems only survive common-envelope evolution if $\alpha_{CE}
\gtrsim 0.5$.) Small kick velocities therefore suppress birth rates
(Figure 11), most severely among short-period systems (Figure 12). 

{\em Mass ratio distribution}.  As noted above, the range in primary
masses ($\sim 15-25$\,M$_\odot$), secondary masses ($\sim
0.5-1.5$\,M$_\odot$)  and orbital periods ($\sim 2-5$\,yr) from which
progenitor binaries are drawn (see Figure 4) is far beyond exploration by
current observational techniques. We consider that our adopted mass ratio
distribution represents a plausible maximum frequency to such systems,
consistent with constraints of dynamical stability.  The birth rates we
derive must therefore be considered upper limits.  Alternative choices of
mass ratio distribution produce lower birth rates; to the extent that
they differ greatly in function form within the mass ratio window of
interest ($q\sim 0.04-0.1$), they may also alter the character of the
LMXB distribution with respect to structural parameters. For example,
Figure 13 illustrates the period distribution derived for a mass ratio
distribution which is independent of $q$ (apart from a very weak
dependence introduced by retention of the Poisson cutoff parameter) below
a critical mass ratio, $q_c=0.35$. (Such a distribution closely resembles
those used for example by Pols et al.\markcite{P91} 1991, and Dalton \&
Sarazin\markcite{D95} 1995) In this case, the total birth rate decreases
by a factor of $\simeq 27$, and there is a relative shift among surviving
systems from the period range $0.2-0.5$ days to the range $0.5-1$ days, a
consequence of the flattening in mass ratio distribution in the range of
interest, $q\sim 0.04-1$.  For values of $\alpha_{CE}$ close to unity
(not shown), a relative excess of short-period systems appears below
$\sim 0.2$ days, but these systems do not survive common envelope
evolution in our reference case.  Unfortunately, these variations tend to
be confined largely to the short-period ($P_X < 20^{\mbox{h}}$)  part of
the orbital period distribution, where they are easily masked by secular
evolution.  The number ratio of long-period to short-period systems,
which is the principal factor influencing systemic velocities as well, is
only weakly dependent on the distribution of donor stars in mass (cf.\
Figure 6), so long as most of those donors are massive enough ($\gtrsim
1.0$\,M$_\odot$) to evolve to interaction. 

{\em Maximum neutron star mass}.  Given our observationally-motivated
assumption that neutron stars are born with uniform gravitational masses of
1.4\,M$_\odot$, this factor enters only into the estimate of the fraction of LMXB
accretors which may evolve to collapse to a black hole.  Estimates of this
fraction for a range of equations of state (Cook et al.\,1994), along with
the observational limit (Table 1) demand that the equation of state be
relatively stiff and the maximum baryonic mass for neutron stars exceed $\sim
1.9$\,M$_\odot$.

\section{CONCLUSIONS}

On undertaking this study, we hoped that the population synthesis
calculations described here would identify some feature or features
among observable parameters of LMXBs which might be unique artifacts of
their primordial distribution and of the evolutionary pathways leading
to the LMXB state. The analytic technique we have used to execute our
synthesis calculations offers enormous advantages for this purpose over
Monte Carlo approaches, as it is free of statistical noise, and can in
principle yield arbitrarily high resolution in the distribution of final
parameters (or of intermediate parameters), should it be warranted, at
minimal additional computational cost.  Our initial hopes have been
confounded by the realization that supernova kicks must play a pivotal
role in the formation of LMXBs, one which severely limits our ability to
probe their origins on the basis of their observed properties.  We see
three important conclusions emerging from this study: 

{\bf (1)  In the absence of supernova kicks, no LMXBs are formed at
short (${\boldmath P_X \lesssim 1^{\mbox{d}}}$)  orbital periods.}
Stellar winds from the helium star component during the
post-common-envelope, pre-supernova phase are capable of removing enough
mass to reduce many pre-supernova systems to less than twice the mass of
the post-supernova remnant (companion plus neutron star), a necessary
condition for the binary condition to remain bound under instantaneous
mass loss.  However, short-period systems cannot then accommodate the
much greater pre-supernova expansion of the low-mass helium star. 
Unless moderately large natal kicks are imparted to neutron stars (i.e.,
kicks averaging a substantial fraction of the relative orbital velocity
of the binary at the supernova stage), only sufficiently long-period
systems survive, and then only if $\alpha_{CE}$ is large ($\alpha_{CE} >
0.6$).  These long-period systems all contain giant branch donors, and
transfer mass at super-Eddington rates. 

This conclusion in fact applies not only to the evolutionary channel
explored here, but to any putative formation channel in which the
neutron star progenitor has a non-degenerate envelope.  Stars with
massive degenerate cores and hydrogen-rich envelopes, either in place of
or in addition to helium envelopes, become red supergiants, and could
leave only extremely long-period neutron star binaries. Only
accretion-induced collapse, in which the neutron star progenitor is
virtually completely degenerate, could allow pre-SN systems close enough
(and with little enough gravitational mass lost in the collapse)  to
produce short-period LMXBs in the absence of supernova kicks.  However,
whether accretion-induced collapse is a viable neutron star formation
mechanism remains an unresolved issue:  We are not aware of any
plausible model which would feed accreted matter through a
hydrogen-burning shell fast enough to stabilize helium burning (and
thereby avoid mass loss during helium runaways) on a massive degenerate
core; on the contrary, evolutionary models of luminous asymptotic giant
branch stars invariably display thermally-pulsing helium shells (Iben \&
Renzini\markcite{I83} 1983).

{\bf (2)  The characteristics of newborn LMXBs are almost entirely
independent of the history of their progenitors.} The ranges in donor
masses and orbital periods allowed to LMXBs are dictated by age and
stability constraints at the onset of the mass transfer phase.  The
distribution of systems over these parameters is influenced primarily by
the efficiency of magnetic braking, which separates short- from
long-period LMXBs.  To a much smaller extent, it is also affected (i) by
the average magnitude of the supernova kick, the effect being more
evident when this average tends to very small values (i.e.,
disappearance of short-period LMXBs in the absence of kicks); and (ii)
by the common envelope efficiency, values of $\alpha_{CE} < 0.1$
precluding LMXB formation altogether.  Apart from these extreme
circumstances, supernova kicks obliterate any memory of how binaries
arrived at the supernova stage;  the LMXB distribution carries virtually
no information about their evolutionary history.  As a result,
alternative formation mechanisms are indistinguishable, except where an
evolutionary channel leads to pre-SN binaries dramatically different
from those relevant to the present study, e.g., the direct-SN mechanism
(Kalogera 1997).  Common envelope evolution, which
characterizes all other LMXB formation channels proposed to date,
inevitably leads to similar distributions of short-period pre-SN
binaries, sharing as their most prominent feature a short-period cutoff
dictated by the dimensions of donor and pre-SN components. 

{\bf (3)  Except as upper limits, theoretical estimates of galactic LMXB
birth rates are not credible.} These estimates depend one-for-one on the
birth frequency of primordial binaries with suitable initial properties
(in our case, $M_1 \sim 12 - 25$\,M$_\odot$, $M_2 \sim 0.5 -
2$\,M$_\odot$, and $P \sim 2 - 5$\,yr ($A \sim 800 -
1800$\,R$_\odot$).  While details may vary somewhat, all LMXB formation
channels (including those proceeding through accretion-induced collapse)
appeal to a primordial population of massive stars ($M_1 \gtrsim
10$\,M$_\odot$)  with low-mass companions ($M_2 \lesssim
2$\,M$_\odot$) in long-period orbits ($P > 1$\,yr). The true frequency
of such systems is observationally indeterminate, and constrained in the
number density of low-mass companions a massive star may retain
consistent with dynamical stability.  In our case, we have pushed the
binary frequency to this limit, and so treat our birth rate estimates as
upper limits.  We have found, moreover, that even variations among
possible mass ratio distributions within the range of interest are
probably obscured in their effect on LMXB properties by secular
evolution in that state. 

Our conclusions regarding the role of supernova kicks in LMXB formation
support and extend those reached independently by Terman, Taam \& Savage
(1996; hereafter TTS96).  In contrast, Iben, Tutukov, \& Yungel'son
(1995;  hereafter ITY95) found such kicks unnecessary.  This difference
appears to have its origin in several factors.  One is the definition of
common-envelope efficiency.  That which we use is identical with that
employed by TTS96;  as previously noted by Han, Podsiadlowski \&
Eggleton (1995) and again by TTS96, the expression used by ITY95
understates the binding energy of the envelope by a factor of two or
three, whereas detailed numerical simulations presented by Rasio \&
Livio (1996)  are consistent with our expression (eq.\ [A8]).  ITY95
thus find wider post-common envelope systems, capable of accommodating
the radial expansion of the helium star progenitors of neutron stars.
Interestingly, in this regard, their models with assumed efficiency
$\alpha_{CE}= 0.5$, corresponding roughly to our $\alpha_{CE} = 1$,
produce no LMXBs with main sequence donors (see Table 1 in ITY95), in
agreement with our results.  A second major difference concerns the
extent and consequences of wind mass loss from helium stars.  In
contrast to our models and to TTS96, ITY95 find significant
contributions to the total LMXB birth rate from systems undergoing case
B mass transfer, which leave post-common envelope core helium burning
primaries.  We find that the extensive mass loss suffered by helium
stars during core helium burning (eq.\ [A9]) greatly expands the range
of initial helium star masses and separations for which Roche lobe
overflow will abort evolution prior to core collapse (cf.\ Figure 2,
eq.\ [A10]), eliminating such stars as viable LMXB progenitors.

A final word is on order regarding angular momentum loss rates due to
magnetic braking.  We have not explored the dependence of our results on
variants of our adopted braking rate.  Qualitatively, stronger braking
will enable wider post-supernova systems to form short-period LMXBs. For
example, King \& Kolb\markcite{KK97} (1997) were able to produce
short-period LMXBs with donors more massive than $1.3$\,M$_\odot$ without
invoking kicks (these are not included by ITY95 or TTS96), because they
assume a magnetic braking law stronger than ours by about an order of
magnitude. However, our interpretation of braking rates among single stars
indicates that magnetic braking is strongly suppressed at masses this
large (cf.\ eq.\,[9]). 

\acknowledgments

It is a pleasure to thank an anonymous referee for comments that helped us
improve the focus of our paper. We are grateful to D.\ Psaltis for
numerous valuable discussions and for carefully reading the manuscript. We
also thank I.\ Iben, A.\ V.\ Tutukov, and L.\ R.\ Yungel'son for
stimulating and enlightening discussions, and their help in identifying
the root causes for differences in our results; and U.\ Kolb for help in
resolving the effect of the efficiency of magnetic braking.  This work was
supported by National Science Foundation under grant AST92-18074 and the
Graduate College of the University of Illinois under a Dissertation
Completion Fellowship. 

\appendix
\section{ANALYTIC APPROXIMATIONS USED IN THE MODEL}

Following are the basic analytic relationships employed in our population
synthesis models for the formation of LMXBs. They are grouped in roughly
the sequence in which they enter consideration along the evolutionary path
from primordial binary to ZALMXB. References identify the sources of the
relationships used here, or (for stellar models) the detailed calculations
which we here analytically approximate. The stellar models in each case
assumed solar composition. The units used throughout are: {\em masses}
$(M)$ in M$_\odot$; {\em radii} $(R)$ and {\em orbital separations} $(A)$
in R$_\odot$;  {\em orbital periods} $(P)$ in days; {\em orbital angular
frequencies} $(\omega)$ in $Hz$; and {\em evolutionary times} $(t)$ in
years. Natural logarithms are written ``ln'', decimal logarithms ``log'',
and arguments of trigonometric functions are in radians. 

{\em Massive stars} (Schaller et al. 1992; Woosley \&
Weaver\markcite{W86} 1986):

Total stellar mass, reduced by stellar wind losses, of a star at core
helium ignition, $M_{1,i}$, and at core helium exhaustion $M_{1,e}$, as a
function of its initial mass, $10$\,M$_\odot < M_{1,o} < 40$\,M$_\odot$: 
\begin{equation}
\log M_{1,i}=0.9454\log M_{1,o}+0.0533\hspace{9.cm}
\end{equation}
\begin{eqnarray}
\log M_{1,e} & = & 0.81\log M_{1,o}+0.174 
\hspace{6.5cm} M_{1,o}\leq20\mbox{\,M}_\odot \hspace{1.cm}\nonumber \\
& = & \frac{1}{2}\left[(0.81\log M_{1,o}+0.174)(1-\sin \phi)+
\right.\nonumber \\
& & \left. 0.9095(1+\sin \phi)\right] \hspace{5.2cm} 
20\mbox{\,M}_\odot < M_{1,o} < 40\mbox{\,M}_\odot  
\end{eqnarray}
where $\phi=10(\log M_{1,o}-\log(20)-\pi/20)$.
Mass of the helium core, $M_{He}$, produced by a star of initial mass 
$M_{1,o}$ {\em before} central He ignition:
\begin{equation}
\log M_{He}=1.589\log M_{1,o}-1.393.\hspace{9.cm}
\end{equation}
If the massive star evolves through the core He burning phase, the He-core
mass grows in mass by $\simeq$1.1\,M$_\odot$ because of shell-hydrogen
burning. The helium core is subsequently exposed by common envelope
evolution, becoming the primary component mass in the next evolutionary
phase.

Radii of stars at core helium ignition, $R_{1,i}$, at core helium exhaustion, 
$R_{1,e}$, and at core collapse, $R_{1,SN}$:
\begin{eqnarray} 
\log R_{1,i} & = & 1.0785\log M_{1,o}+1.5123 
\hspace{6.cm} M_{1,o}\leq 20\mbox{\,M}_\odot\hspace{1.cm} \nonumber \\    
& = & \frac{1}{2}\left[(1.0785\log M_{1,o}+1.5123)(1-\sin \phi)+ 
\right.\nonumber \\
& & \left.(1.053\log M_{1,o}+1.111)(1+\sin \phi)\right]\hspace{2.3cm} 
20\mbox{\,M}_\odot < M_{1,o} < 40\mbox{\,M}_\odot   
\end{eqnarray} 
where $\phi=15(\log M_{1,o}-\log(20)-\pi/30)$,
\begin{eqnarray}  
\log R_{1,e} & = & 1.5745\log M_{1,o}+0.97125 \hspace{5.6cm} 
M_{1,o}\leq 20\mbox{\,M}_\odot \hspace{1.cm}\nonumber \\      
& = & \frac{1}{2}\left[(1.5745\log M_{1,o}+0.97125)(1-\sin \phi)+
\right.\nonumber \\ 
& & \left. 0.74(1+\sin \phi)\right] \hspace{5.6cm} 
20\mbox{\,M}_\odot < M_{1,o} < 40\mbox{\,M}_\odot                             
\end{eqnarray}  
where $\phi=12(\log M_{1,o}-\log(20)-\pi/24)$,
\begin{eqnarray}   
\log R_{1,SN} & = & 1.148\log M_{1,o}+1.5888 \hspace{5.6cm} M_{1,o}\leq 
20\mbox{\,M}_\odot \hspace{1.cm}\nonumber \\  
& = & \frac{1}{2}\left[(1.148\log M_{1,o}+1.5888)(1-\sin \phi)+ 
\right.\nonumber \\
& & \left. 0.65(1+\sin \phi)\right]\hspace{5.2cm} 
20\mbox{\,M}_\odot < M_{1,o} < 40\mbox{\,M}_\odot             
\end{eqnarray}   
where $\phi=12(\log M_{1,o}-\log(20)-\pi/24)$. 

{\em Roche geometry} (Eggleton\markcite{E83} 1983):

Dimensionless radius of the Roche lobe of component 1 $(r_{L_1}\equiv 
R_{L_1}/A)$ 
as a function of binary mass ratio $(q_1\equiv M_1/M_2)$: 
\begin{equation}
r_{L_1}=\frac{0.49q_1^{2/3}}{0.6q_1^{2/3}+\ln(1+q_1^{1/3})}.
\end{equation}
Component indices are interchangeable in this expression. 

{\em Common envelope evolution} (Webbink\markcite{W84} 1984):

Ratio of post-common envelope binary separation, $A_f$, to pre-common envelope 
separation, $A_i$:
\begin{equation}
\frac{A_f}{A_i} = \frac{\alpha_{CE}r_{L_1}}{2}\left(\frac{M_2}{M_1}\right)
\left(\frac{M_{He}}{(M_1-M_{He})+\frac{1}{2}\alpha_{CE}r_{L_1}M_2}\right). 
\end{equation}

{\em Helium stars} (Habets 1985; Woosley, Langer, \& Weaver 1995):

Helium stars experience mass loss in a wind and their masses can decrease
significantly during the central-He burning phase. The 
final mass of a helium star, $M_{He,f}$, at supernova as a function of its
mass, 
$M_{He}$ at core helium ignition is approximated by:
\begin{equation}
M_{He,f} = 3.64-6.42\mbox{exp}\left[-\frac{(M_{He}-3.43)^{0.33}}{0.55}\right]
~~~~4 \mbox{\,M}_\odot < M_{He} < 20 \mbox{\,M}_\odot
\end{equation}
If the helium star is exposed after central He exhaustion then it is not
affected by mass loss and its mass at supernova is equal to its mass at
the end of the time of its exposure. 

Radius of helium star at supernova,
$R_{He,f}$: \begin{eqnarray} 
R_{He,f} & = & 3.0965-2.013\log M_{He,f}~~\hspace{3.2cm}
M_{He,f}\leq 2.5\mbox{\,M}_\odot \nonumber \\
& = & 0.0557\left[(\log M_{He,f}-0.172)^{-2.5}\right] 
~~~~~~~~~~~~~~~~M_{He,f}>2.5 \mbox{\,M}_\odot
\end{eqnarray}

{\em Angular momentum loss}:

Loss rate from gravitational radiation for a circular orbit 
(Landau \& Lifshitz\markcite{L51} 1951):
\begin{equation}
\dot{J}_{GR} = - \frac{32}{5}\frac{G}{c^5}~
\left(\frac{M_{NS}M_2}{M_{NS}+M_2}
\right)^2~A^4~\omega^5,
\end{equation}
where $G$ is the gravitational constant, $c$ is the speed of light, 
and $\omega$ 
is the orbital frequence. We neglect 
the enhancement of gravitational radiation losses in eccentric orbits 
(Peters \& Mathews\markcite{P63} 1963). 
The above equation can be integrated over a time interval 
$\Delta t$ required for a circular orbit to decay from orbital period $P_i$ to 
$P_f$:
\begin{equation}
P_f^{8/3} - P_i^{8/3} + 8~A_{GR}~\Delta t = 0,
\end{equation}
where
\begin{eqnarray}
A_{GR} & = & \frac{q~(1+q)^{-1/3}~M_{NS}^{5/3}}{3.75\times 10^{11}} ~
\mbox{yr}^{-1}~\mbox{day}^{8/3} \nonumber \\
q & = & \frac{M_2}{M_{NS}} \nonumber 
\end{eqnarray}

Loss rate from the magnetic stellar wind of a synchronously-rotating secondary 
(cf. Rappaport, Verbunt, \& Joss 1983):
\begin{equation}
\dot{J}_{MB}  =  - 1.8\times 10^{47}~b(M_2)
~M_2~R_2^2~\omega^3,
\end{equation}
where $\dot{J}_{MB}$ is in cgs units (dyne\,cm\,s$^{-1}$) and $b(M_2)$ is the 
magnetic braking efficiency (eq.\,[9]), which becomes equal to zero for fully 
convective stars ($M_2\leq 0.37$\,M$_\odot$). 
For stars with radiative cores ($M_2 > 0.37$\,M$_\odot$),
we neglect the evolutionary expansion of the secondary with time and find 
that during a time interval $\Delta t$ a circular orbit decays from 
orbital period $P_i$ to $P_f$: 
\begin{eqnarray}
& & \frac{a}{4}\left(P_f^{8/3}-P_i^{8/3}\right)-
\frac{a^2}{3}\left(P_f^{2}-P_i^{2}\right)+
\frac{a^3}{2}\left(P_f^{4/3}-P_i^{4/3}\right)
-a^4\left(P_f^{2/3}-P_i^{2/3}\right) \nonumber \\
& & \hspace{4.cm}+a^5\ln \frac{1+P_f^{2/3}/a}
{1+P_i^{2/3}/a}+2A_{MB}\Delta t =0,
\end{eqnarray}
where
\begin{eqnarray}
A_{MB} & = & b(M_2)\frac{q^2~(1+q)^{1/3}~M_{NS}^{4/3}}{5.78\times 10^9}
~\mbox{yr}^{-1}~\mbox{day}^{10/3}~, \nonumber \\
a & = & \frac{A_{MB}}{A_{GR}} \nonumber 
\end{eqnarray}

{\em Low-mass stars}:

Radii at ZAMS, terminal main sequence, and at the base of the giant branch, 
along with the time evolution of the stellar radius have been given 
Paper I.

\newpage

\figcaption{Mass as a function of radius for a mass-losing star of initial
mass equal to $15$ M$_\odot$. Open circles indicate the phase of
core-helium burning (after Schaller et al. 1992).}

\figcaption{Limits on orbital separation and primary mass after the
common-envelope ejection for a $1$\,M$_\odot$ secondary and
$\alpha_{CE}=1$.  Thick and thin lines correspond to upper and lower
limits, respectively.  {\it Thick solid line} : first Roche-lobe overflow
just prior to supernova;  {\it solid line} :  first Roche-lobe overflow
just after core-helium exhaustion;  {\it dotted line} :  He-star with a
C-O core (and {\it short-dashed line} : secondary)  accommodated in the
post-CE orbit;  {\it thick long-short-dashed line} :  first Roche-lobe
overflow just prior to core-helium ignition; {\it long-short-dashed line}
: He-star with a helium core accommodated in the post-CE orbit. It is
evident that a non-vanishing area of the parameter space is available to
LMXB-progenitors only if first Roche-lobe overflow occurs after
core-helium exhaustion.}

\figcaption{Limits on orbital separation and primary mass after the
common-envelope ejection for a $1$\,M$_\odot$ secondary and
$\alpha_{CE}=1$, in the case of a {\em symmetric} core collapse. 
Line-type coding is the same as in Figure 2. In addition, {\it thick
dot-short-dashed line} : mass transfer in the post-SN binary is initiated
within $10^{10}$\,yr; {\it thick dot-long-dashed line} : maximum He-star
mass for keeping the post-SN system bound. It is evident that no parameter
space is available to LMXB-progenitors.}

\figcaption{Limits on orbital separation and primary mass of primordial
(O,B)  and post-common envelope binaries with a 1\,M$_\odot$ secondary,
for $\alpha_{CE} = 1$. Line-type coding is the same as in Figure 2.}

\figcaption{Distribution of primordial binaries with primary mass
$M_1=20$\,M$_\odot$ and orbital separation $A=1000$\,R$_\odot$ over mass
ratio, $q$. The corresponding secondary masses, $M_2$, are also shown.}

\figcaption{Distribution of nascent LMXBs, $\Phi_P(\log M_2, \log P_X)$,
over donor mass, $M_2$, and orbital period, $P_{X}$, for $\langle 
V_k^2\rangle ^{1/2}=300$\,km\,s$^{-1}$ and $\alpha_{CE}=0.3$. Mass
transfer at {\bf (a)} sub-Eddington rates, and {\bf (b)} both sub- and
super-Eddington rates.}

\figcaption{Limits on donor masses, $M_2$, and orbital periods, $P_{X}$,
of binaries at the onset of mass transfer for a population of age
$10^{10}$\,yr.  Heavy solid lines mark the loci of zero-age main sequence
stars (ZAMS), terminal main sequence stars (TMS), and stars at the base of
the giant branch (BGB). {\it Dot-dashed line}: maximum orbital periods for
mass transfer in a Galactic disk population of age $10^{10}$\,yr; {\it
thin solid lines}: maximum donor masses for thermal stability on the main
sequence and in the Hertzsprung gap, assuming conservative mass transfer;
{\it dotted lines}: maximum donor masses for thermal stability on the main
sequence and in the Hertzsprung gap, and for dynamical stability on the
giant branch, all in the limit that all mass lost from the donor is also
lost from the binary; {\it short-dashed lines}: minimum donor masses for
the development of a delayed dynamical instability; {\it long-dashed
lines}: maximum donor masses for regaining thermal equilibrium after an
initial mass transfer phase on a thermal time scale. }

\figcaption{Distribution of nascent LMXBs, $\Psi_P(\log P_X)$, over
orbital period, $P_{X}$, for $\langle V_k^2\rangle
^{1/2}=300$\,km\,s$^{-1}$ and $\alpha_{CE}=0.3$. {\it Solid line}: both
sub- and super-Eddington systems, and {\it dotted line}: only
sub-Eddington systems.}

\figcaption{Total birth rate of sub-Eddington only (open circles) and sub-
and super-Eddington combined (filled circles) nascent LMXBs as a function
of common-envelope efficiency $\alpha_{CE}$ for $\langle V_k^2\rangle
^{1/2}=300$\,km\,s$^{-1}$.}

\figcaption{Distribution of systems, $\Psi_P(\log P_X)$, transferring mass
at sub- and super-Eddington rates over orbital period, $P_{X}$, for
different values of the common-envelope efficiency, $\alpha_{CE}$, and for
$\langle V_k^2\rangle ^{1/2}=300$\,km\,s$^{-1}$.}

\figcaption{Total birth rate of sub-Eddington only (open circles) and sub-
and super-Eddington combined (filled circles) nascent LMXBs for
$\alpha_{CE}=1$ and sub- and super-Eddington (filled triangles) systems
for $\alpha_{CE}=0.2$ as a function of $\langle V_k^2\rangle ^{1/2}$.}

\figcaption{Distribution, $\Psi_P(\log P_X)$, of combined sub- and
super-Eddington nascent LMXBs over orbital period, $P_{X}$, for different
values of $\langle V_k^2\rangle ^{1/2}$ and for $\alpha_{CE}=0.3$.}

\figcaption{Distribution of systems, $\Psi_P(\log P_X)$, transferring
mass at both sub- and super-Eddington rates over orbital period, $P_{X}$.
The probability density is normalized to the total birth rate, $3.2\times
10^{-6}$\,yr$^{-1}$ for our reference model ({\it solid line}), and
$1.2\times 10^{-7}$\,yr$^{-1}$ for a model with constant mass-ratio
distribution ({\it dotted line}). For both cases $\langle V_k^2\rangle
^{1/2}=300$\,km\,s$^{-1}$ and $\alpha_{CE}=0.3$.}

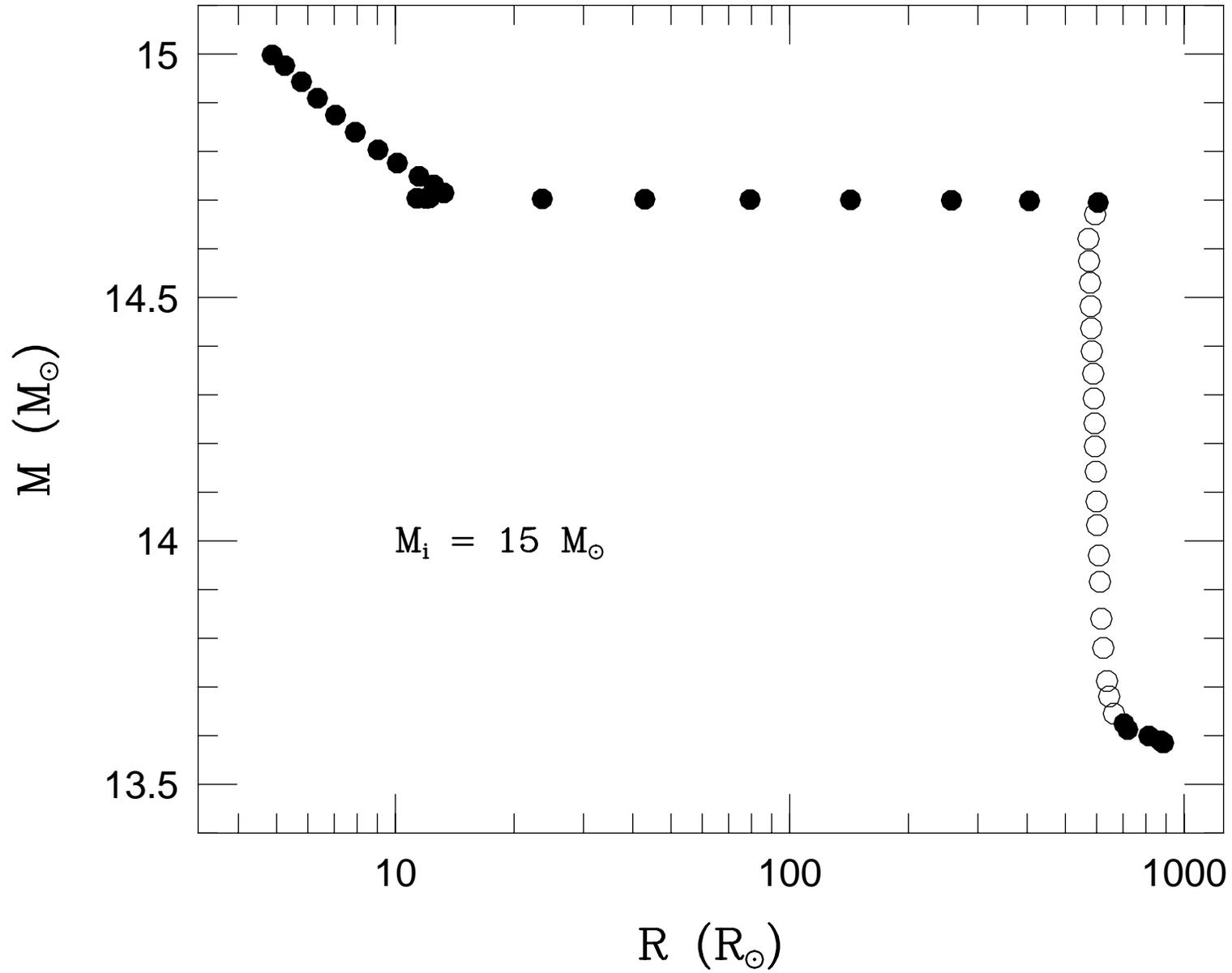
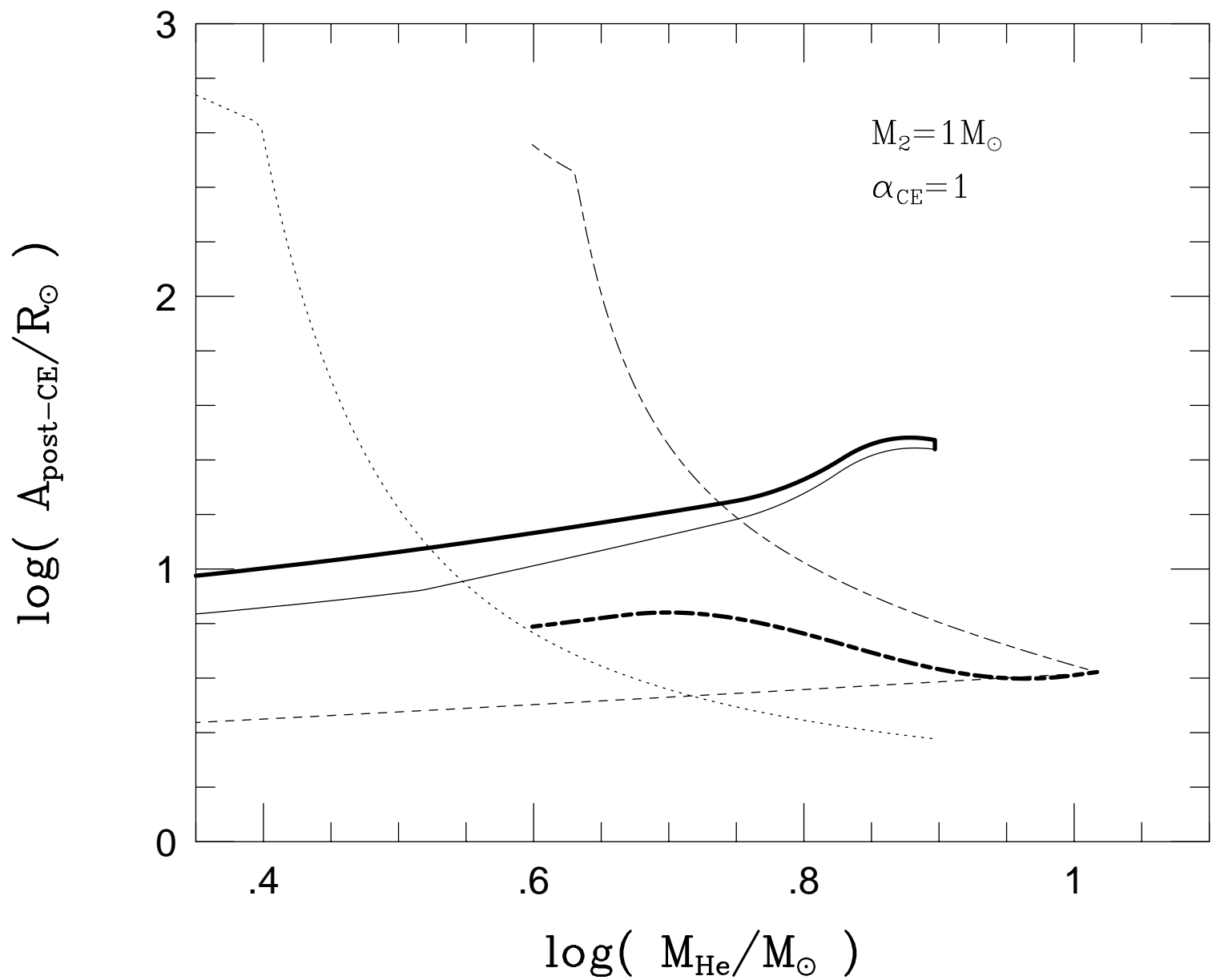
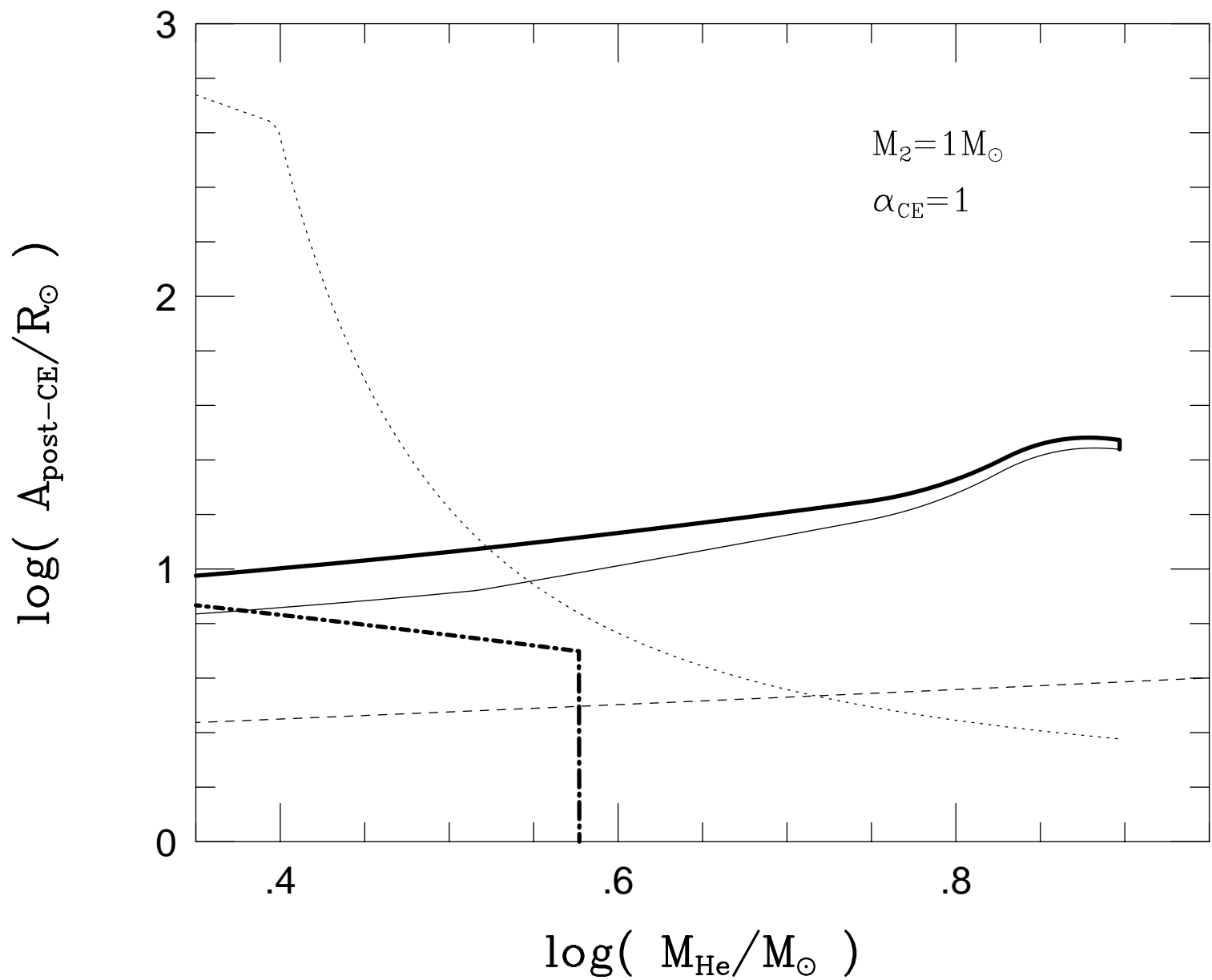
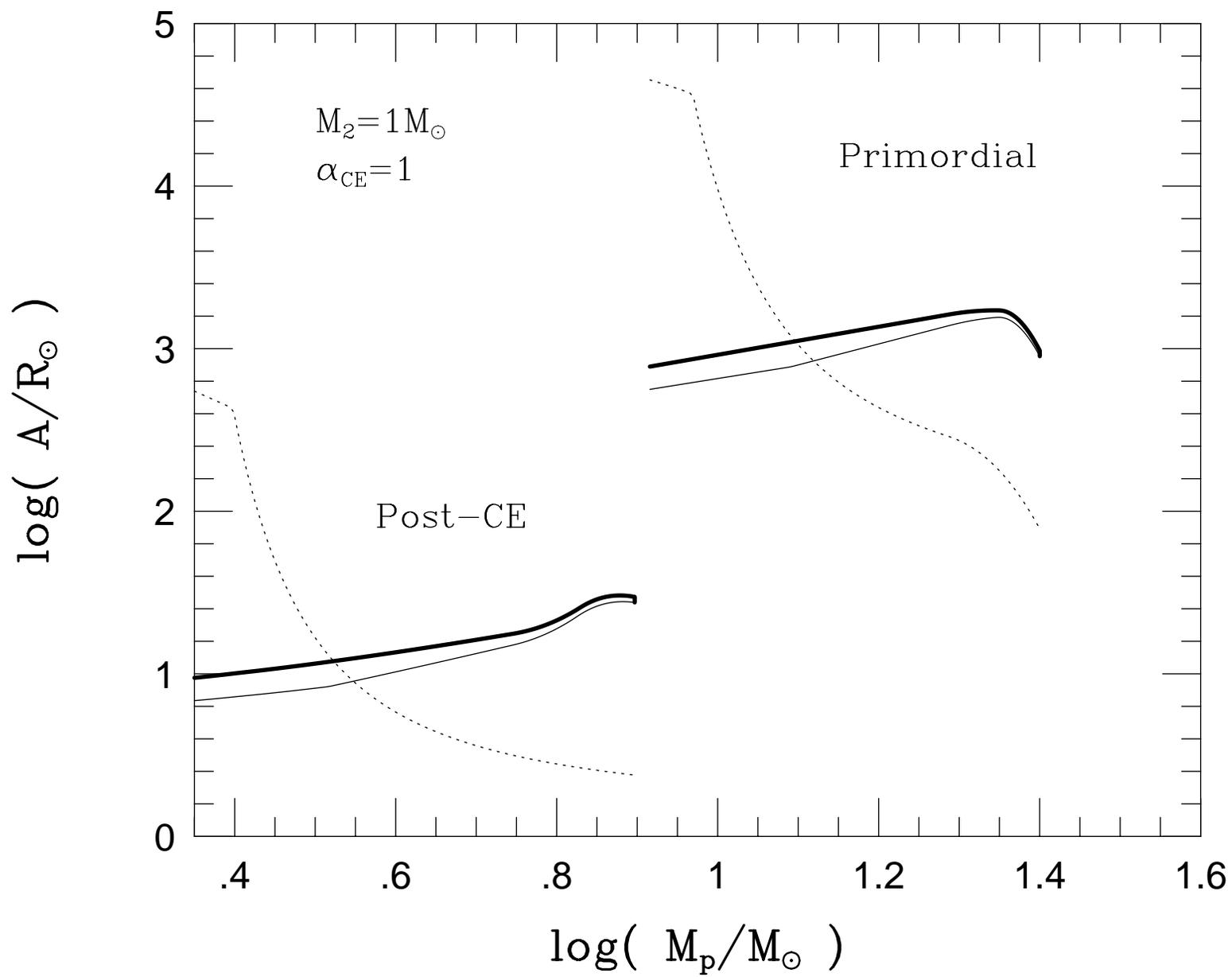
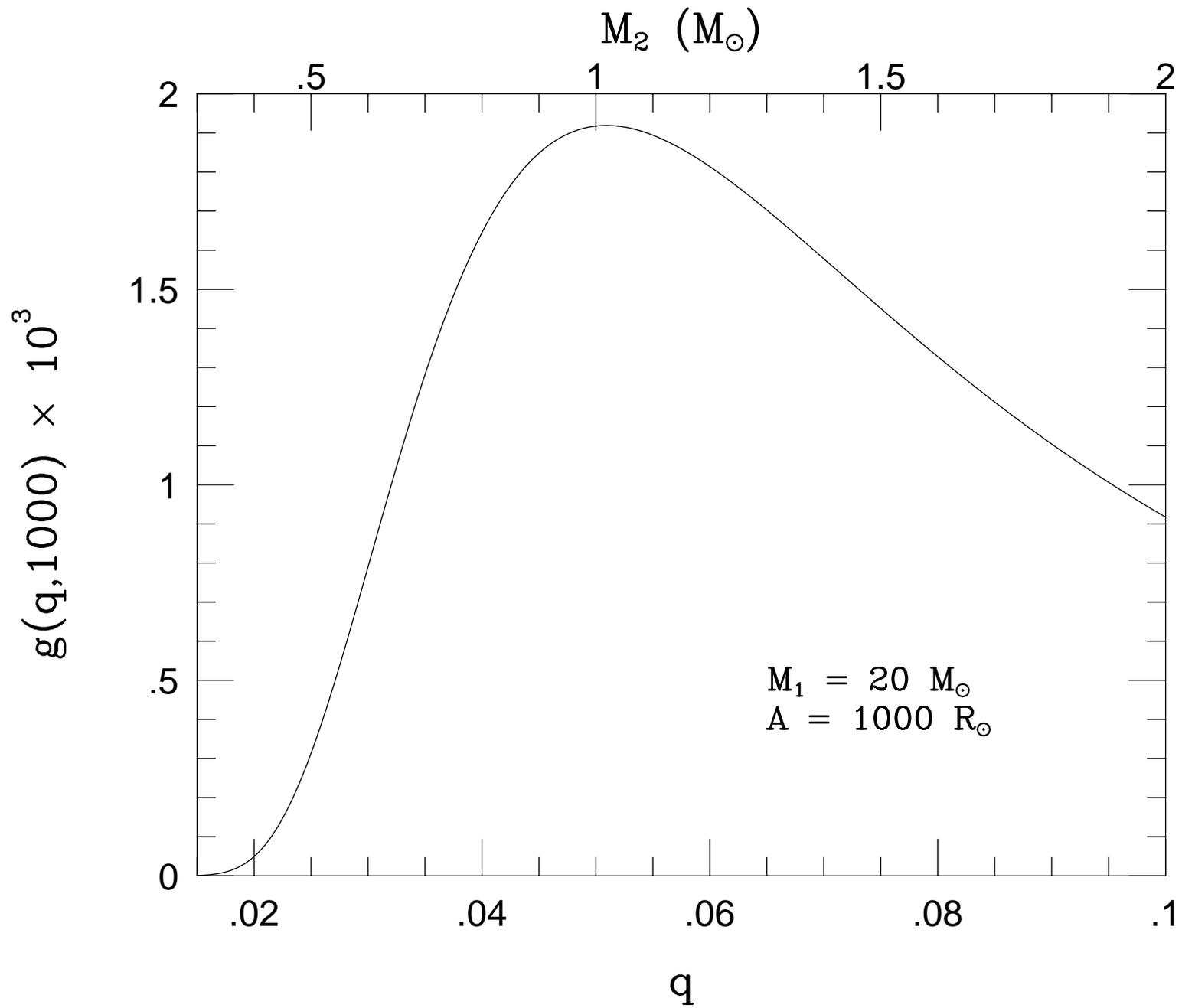
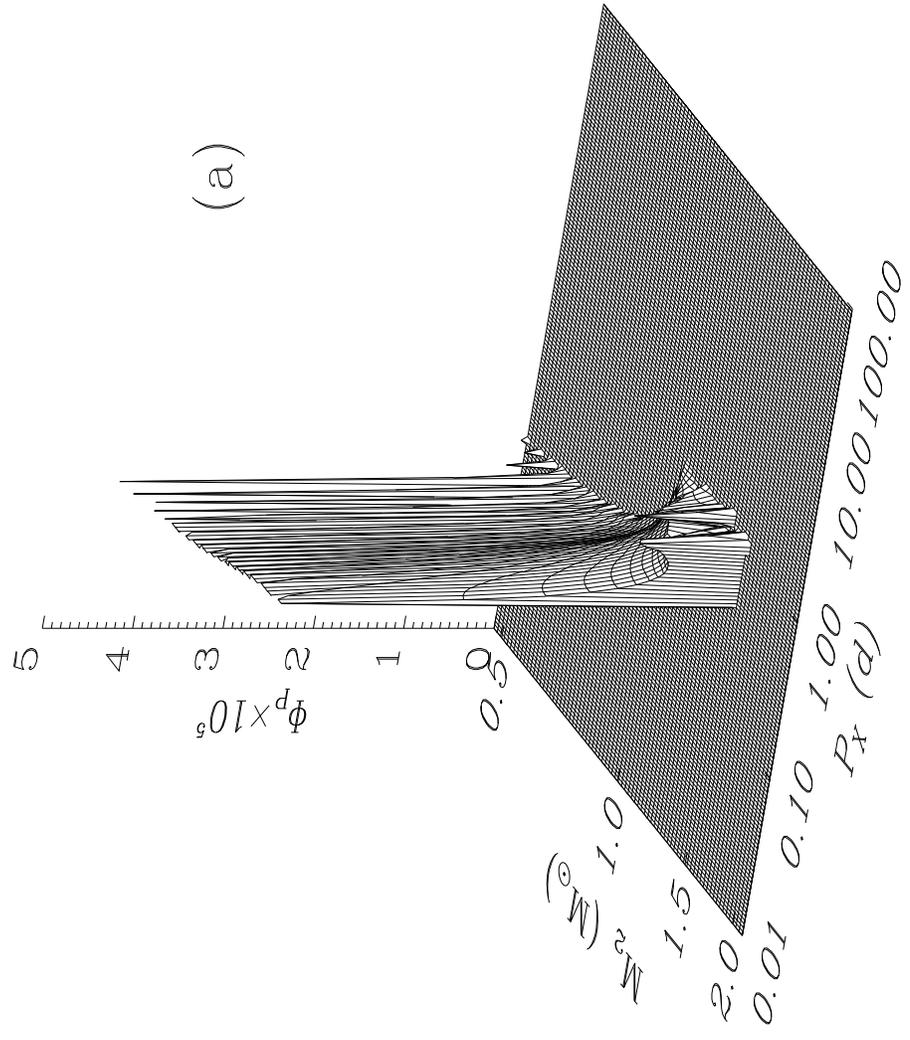
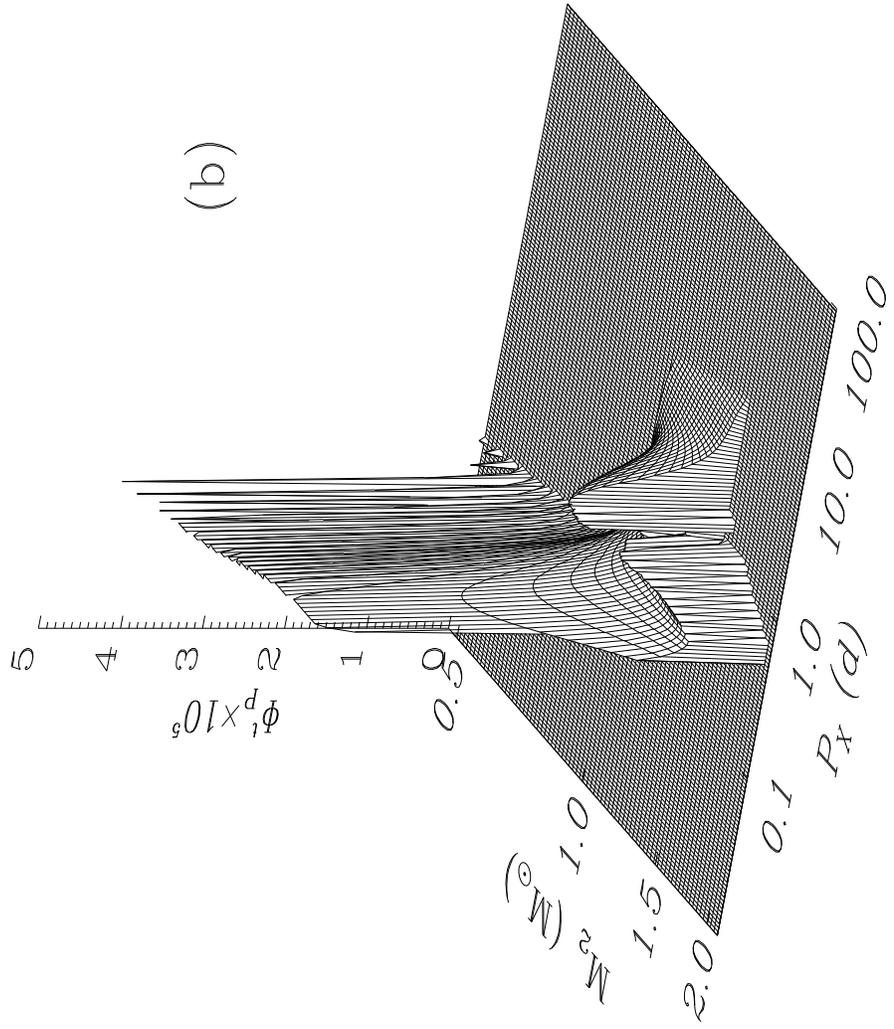
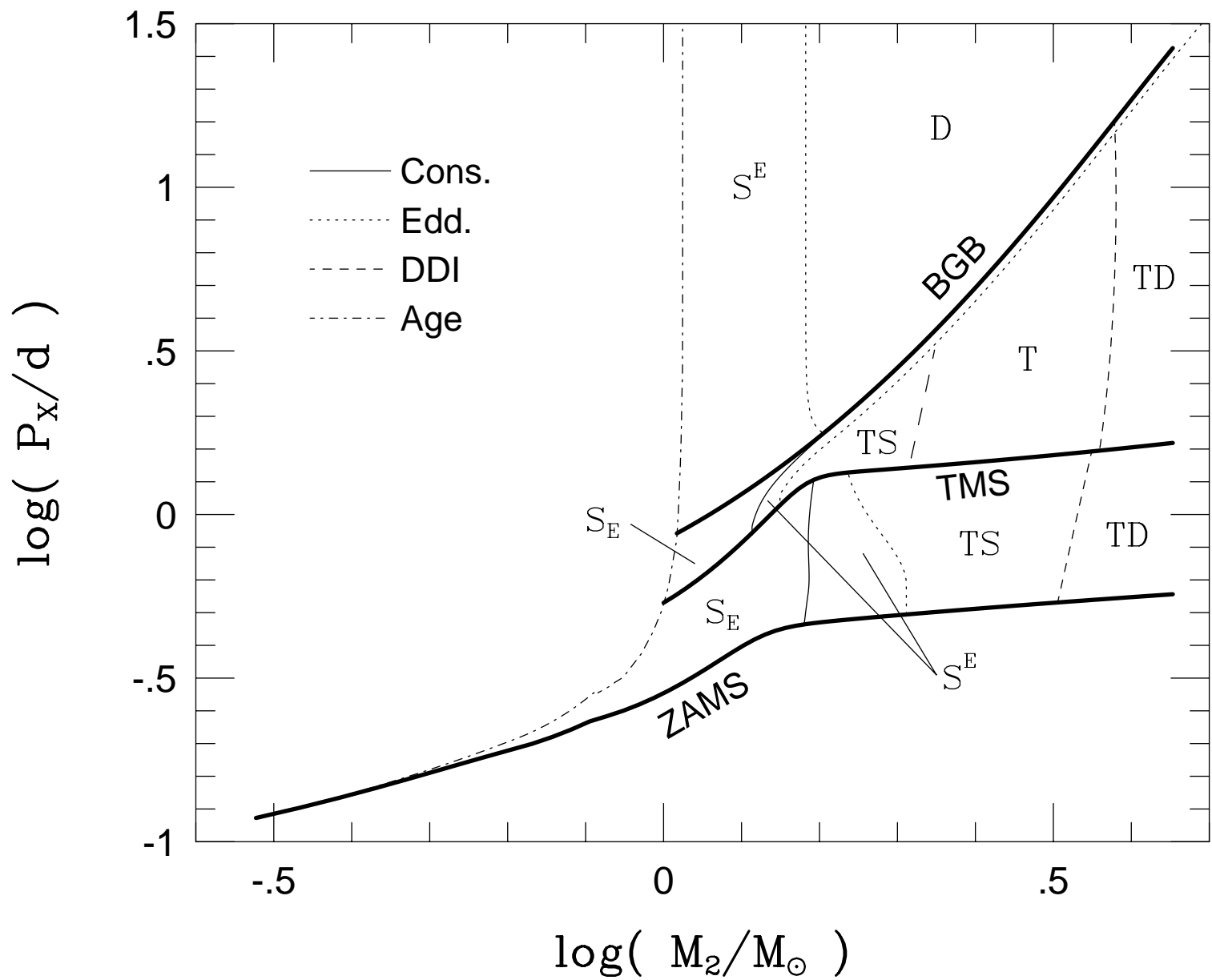
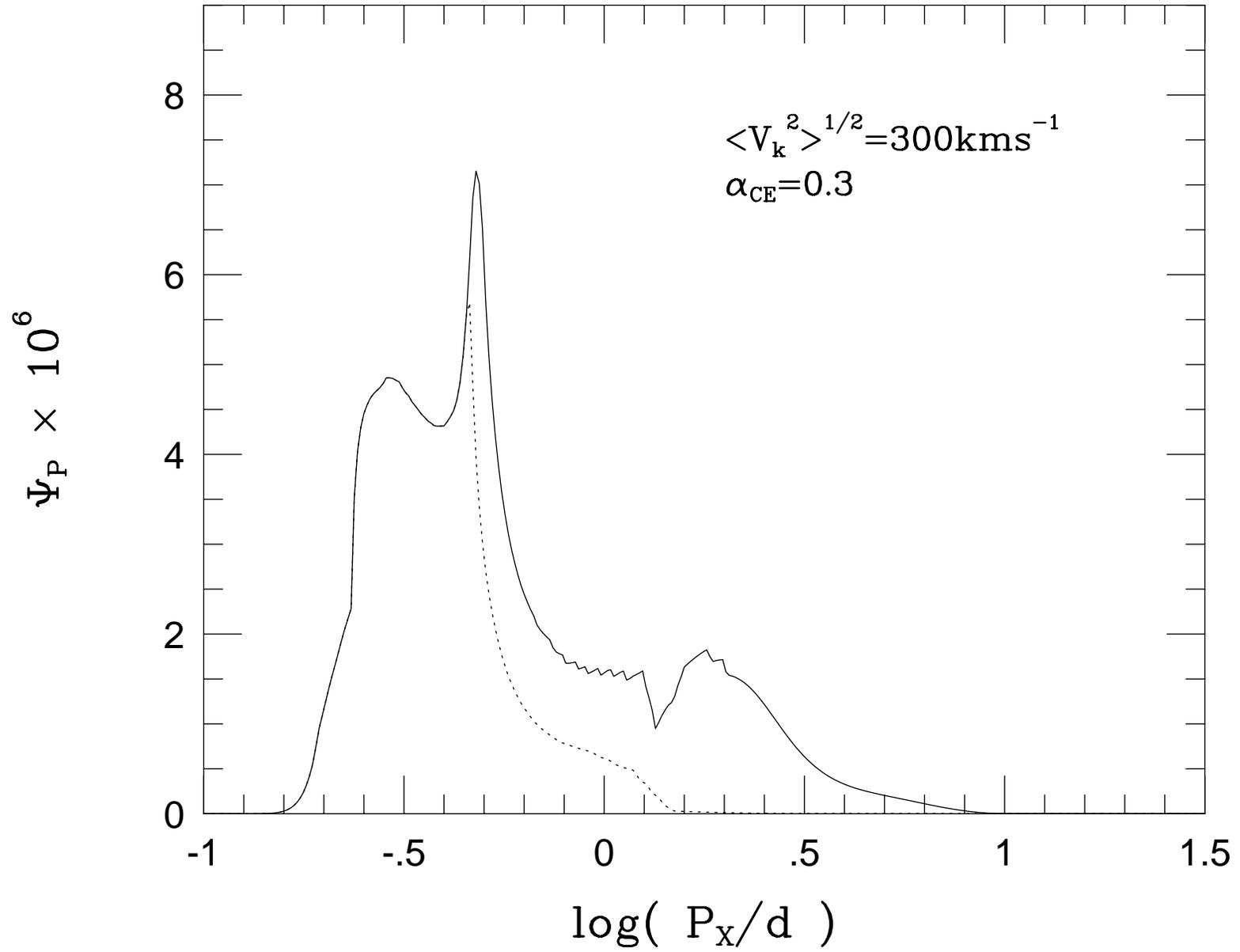
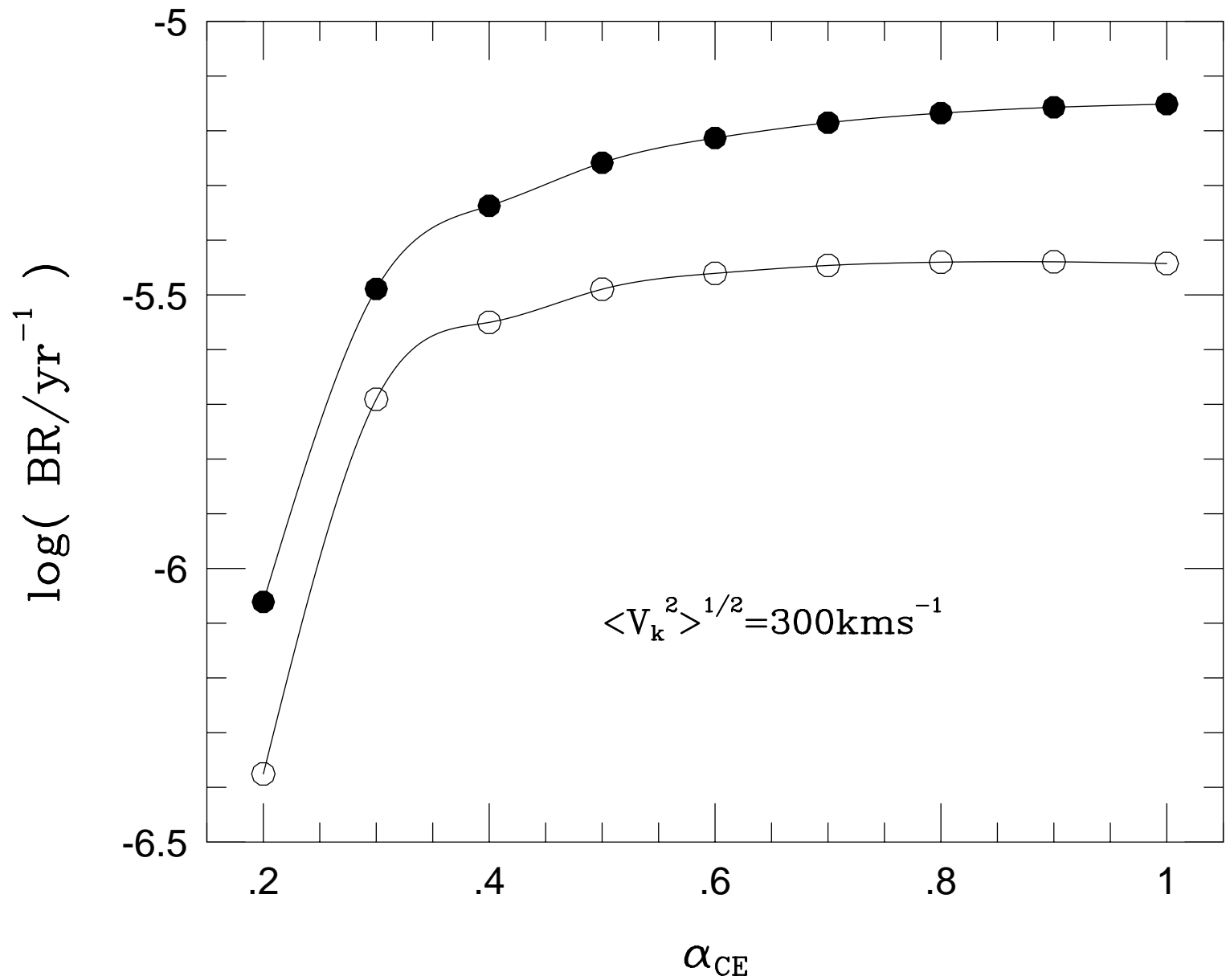
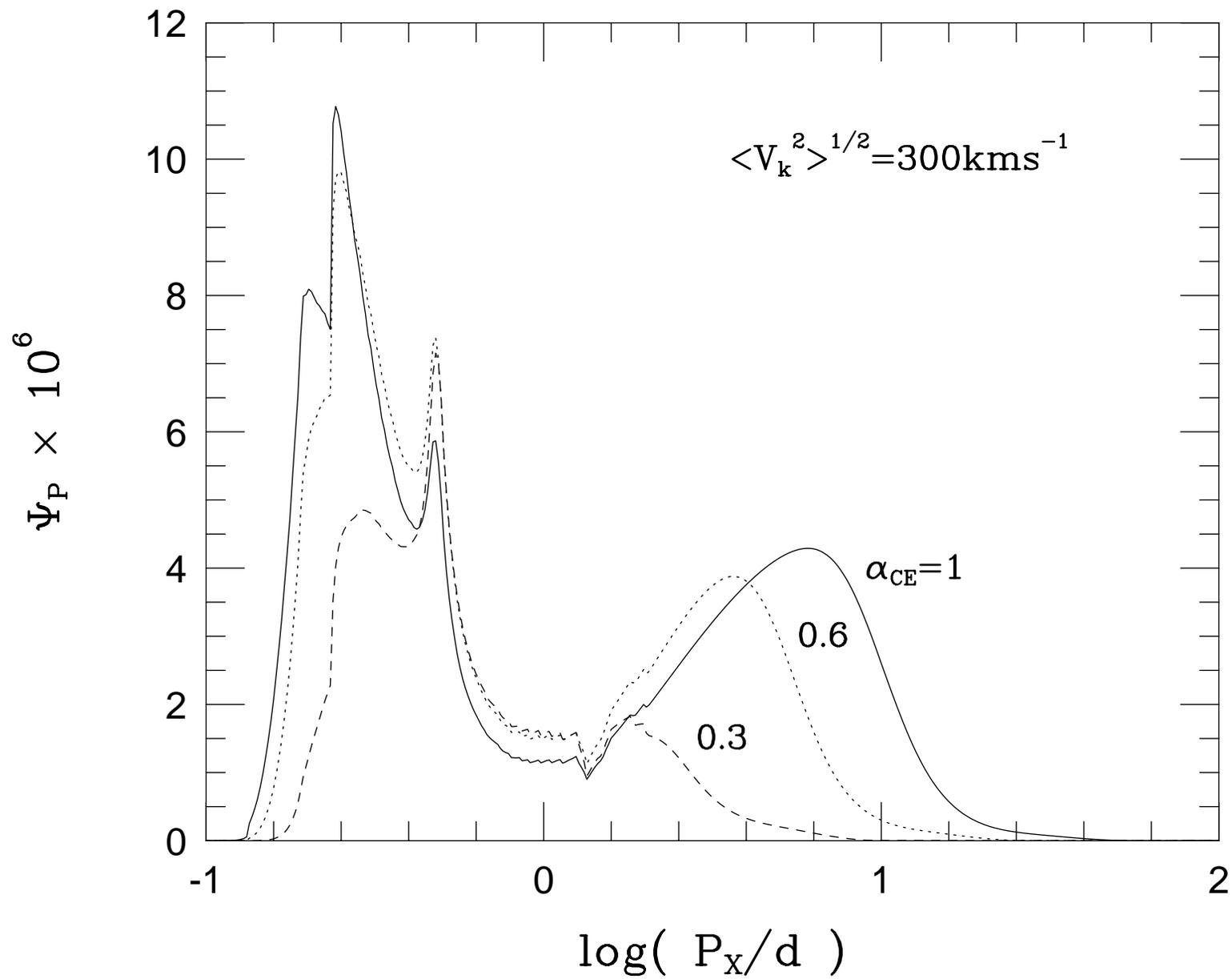
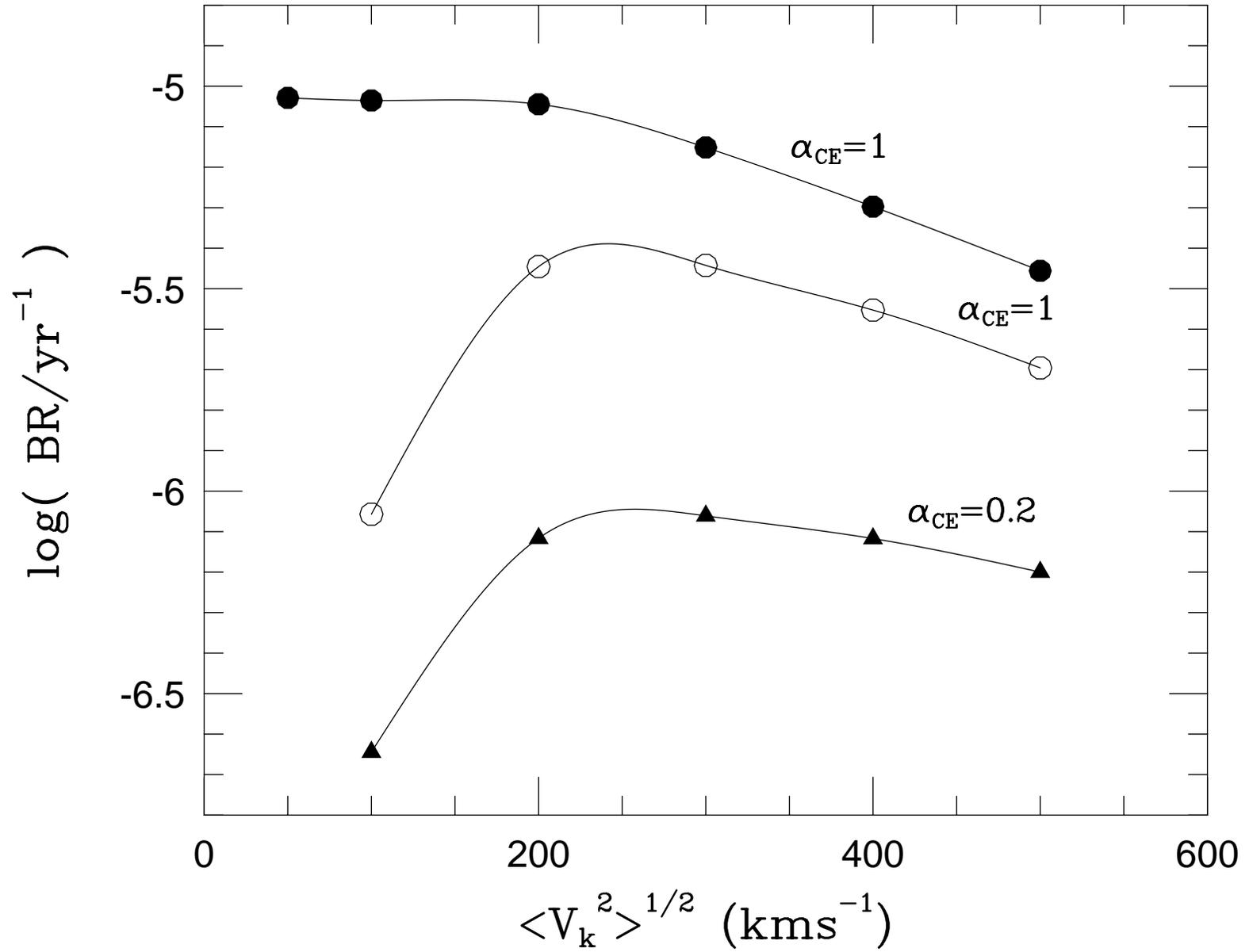
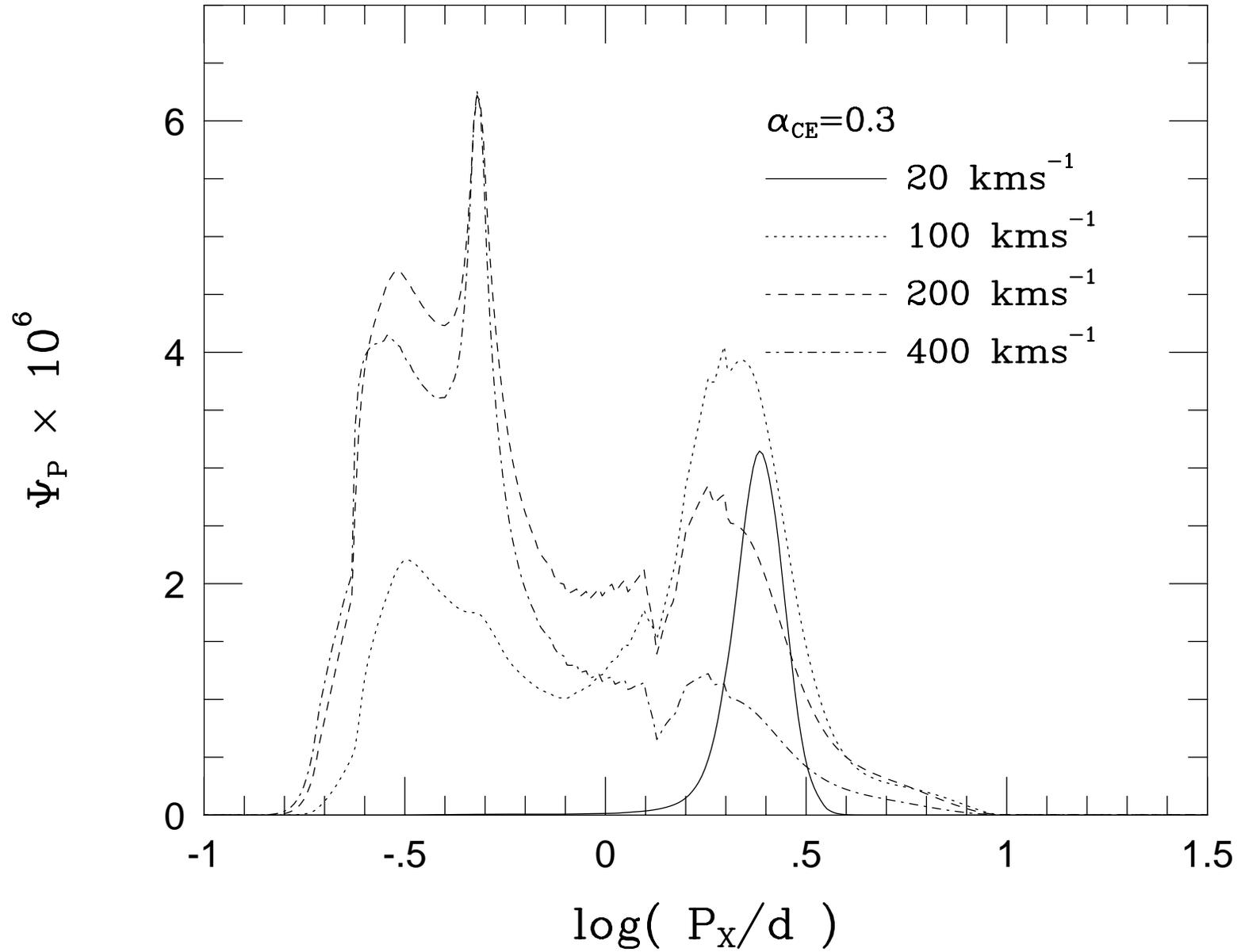
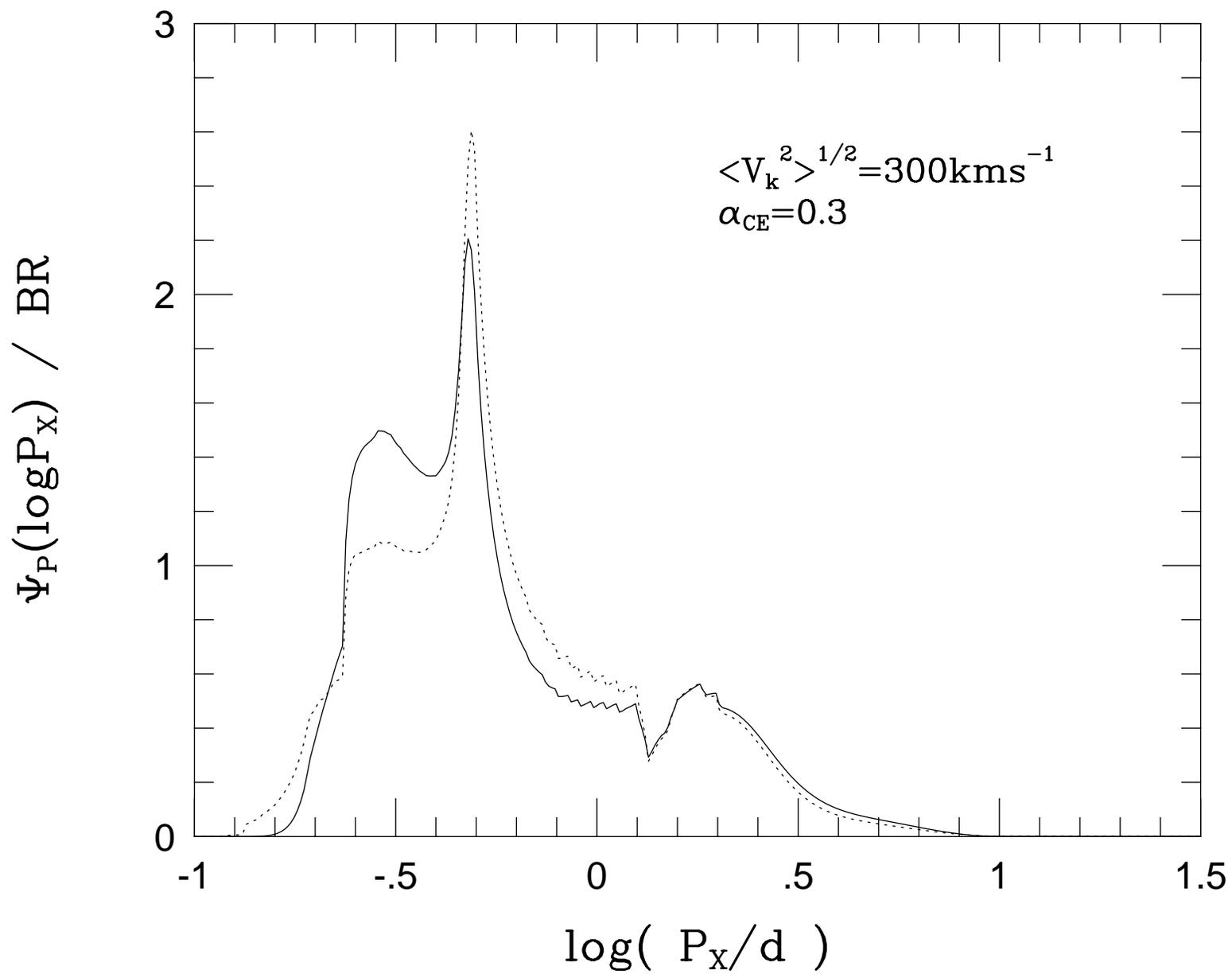
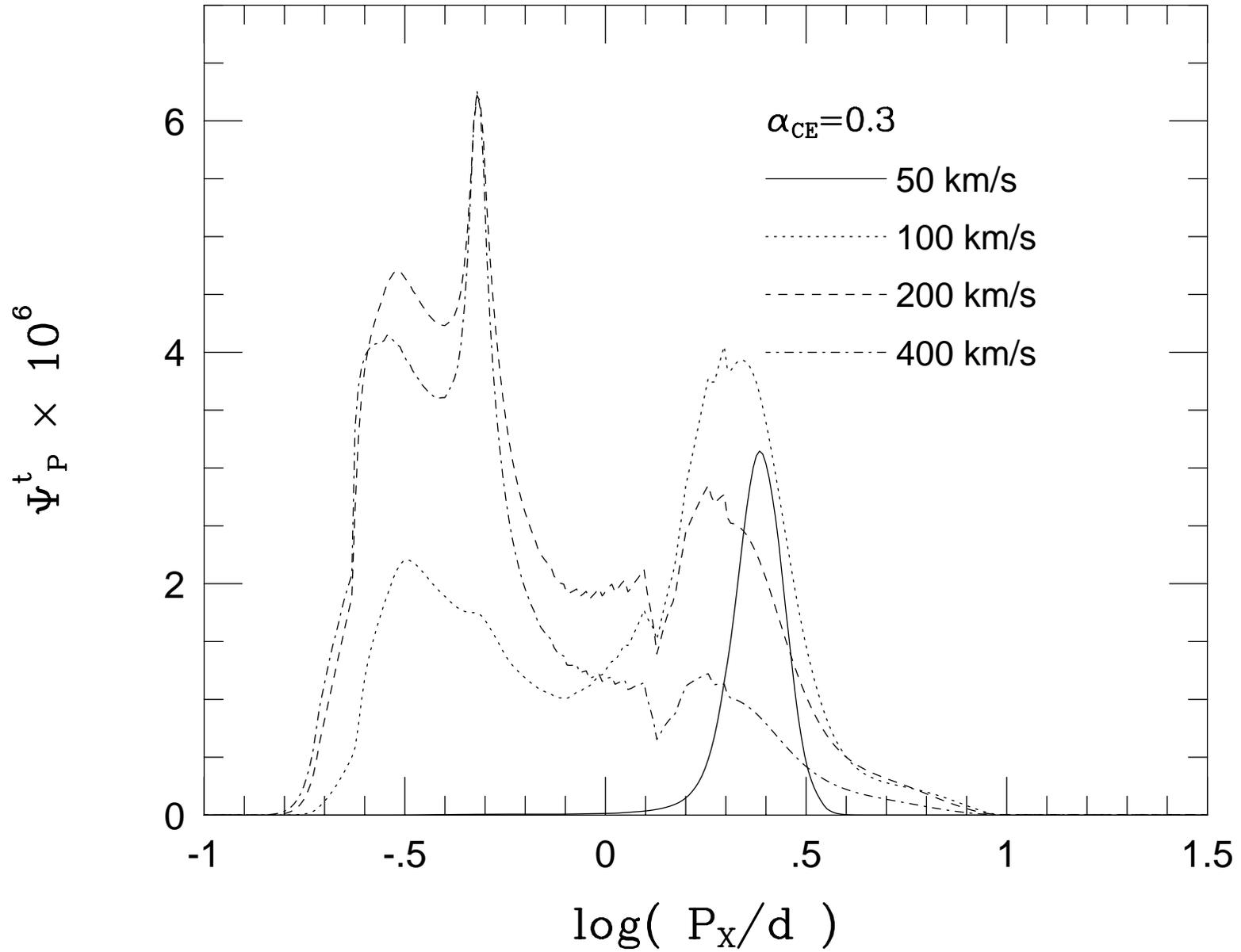
\end{document}